\documentclass[prl,aps,twocolumn,amsmath,amssymb,preprintnumbers,10pt,longbibliography,superscriptaddress]{revtex4-2}

\usepackage{xcolor}
\usepackage{amssymb}
\usepackage{graphicx}
\usepackage{physics}
\usepackage{svg}
\usepackage{bm}
\usepackage[caption=false]{subfig}
\usepackage{siunitx}
\usepackage{enumitem}

\usepackage[page]{appendix}

\renewcommand{\ket}[1]{{|{#1}\rangle}}

\begin{document}
\title{All product eigenstates in Heisenberg models from a graphical construction}

\author{Felix Gerken}
\affiliation{I. Institut für Theoretische Physik, Universität Hamburg,
Germany}
\affiliation{The Hamburg Centre for Ultrafast Imaging, 
Hamburg, Germany}

\author{Ingo Runkel}
\affiliation{Fachbereich Mathematik, Universität Hamburg,
Germany}

\author{Christoph Schweigert}
\affiliation{Fachbereich Mathematik, Universität Hamburg,
Germany}

\author{Thore Posske}
\affiliation{I. Institut für Theoretische Physik, Universität Hamburg,
Germany}
\affiliation{The Hamburg Centre for Ultrafast Imaging, 
Hamburg, Germany}

\date{\today}

\begin{abstract}
Recently, large degeneracy based on product eigenstates has been found in spin ladders, kagome-like lattices, and motif magnetism, connected to spin liquids, anyonic phases, and quantum scars.
We unify these systems by a complete classification of product eigenstates of Heisenberg XXZ Hamiltonians with Dzyaloshinskii-Moriya interaction on general graphs 
in the form of Kirchhoff rules for spin supercurrents.
By this, we construct spin systems with extensive degree of degeneracy linked to exotic condensates which could be studied in atomic gases and quantum spin lattices.
\end{abstract}

\maketitle

In recent years, spin systems were realized in a multitude of physical platforms, including ultra cold atom systems like quantum gas microscopes \cite{Gross2017}, Rydberg atoms \cite{Chen2023,Sbierski2023}, neutral atom arrays with optical tweezers \cite{Ebadi2021,Semeghini2021}, and others \cite{Jepsen2022} as well as in qubit lattices, quantum computers \cite{King2018,Tacchino2020,VanDyke2022} and chains of magnetic atoms on surfaces \cite{Khajetoorians2011,Khajetoorians2013,Chen2022}.
The eigenstates of spin models with Heisenberg exchange interactions, which is an archetypal example for strongly-correlated matter, are highly nontrivial due to their high degree of entanglement, especially of the excited states.
This poses a challenge to both experimental realization and theoretical description.
These models are exactly solvable only in certain limits, e.g., using the Bethe ansatz \cite{Korepin1993}, Jordan-Wigner transformations \cite{Jordan1928,Lieb1961}, or by selecting particular types of interactions \cite{Kitaev2006,Wen2007}.
Contrasting this, 
there recently have been a number of studies on product eigenstates in quantum spin systems that are connected to quantum many-body scars \cite{Moudgalya2022}, which evade the eigenstate thermalization hypothesis \cite{Deutsch1991,Srednicki1994}.

First examples were spin helices in 1D easy-plane quantum magnets \cite{Cao2003,Batista2009,Batista2012,Cerezo2016,Popkov2021,Zhang2021-1,Zhang2021-2,Zhang2022,Ma2022,Popkov2023,Kuehn2023,Shi2023} that have been dubbed phantom helices \cite{Popkov2021,Zhang2021-1,Zhang2021-2}.
These states belong to a chiral reformulation of the Bethe ansatz at fine-tuned parameters of the one-dimensional $XXZ$ quantum Heisenberg model where they consist of zero-energy quasi-particles.
Phantom helices were experimentally observed in ultra cold atoms and generalized to higher dimensions \cite{Jepsen2022}.
Beyond one spatial dimension, product eigenstates can appear as ground states of Heisenberg models with carefully engineered magnetic fields manipulating the individual spins \cite{Cerezo2015,Cerezo2016,Cerezo2017}, 
In the vicinity of a critical point with macroscopic degeneracy, these systems carry a rich phase diagram 
including spin-liquid and broken symmetry phases.
The product eigenstates were studied for their quantum scar properties \cite{Lee2020} and generalized to lattices built from motifs, i.e., elementary building blocks,  
where they are ground states of frustrated lattices at a critical value of their exchange anisotropy \cite{Changlani2018,Pal2021}, also further
showing that product eigenstates exist in quasicrystalline Heisenberg magnets \cite{Chertkov2021}.
Product states are also a starting point for semi-classical and effective approaches to spin systems \cite{Kim2016,Tserkovnyak2020}.
E.g., product states have been used to elucidate the connection between classical and quantum skyrmions \cite{Takashima2016,Sotnikov2022}, 
in particular,
studies indicate that the central spin of quantum skyrmions could decouple in the ground state \cite{Haller2022,Joshi2023}.

In this manuscript, 
we give a procedure to completely classify product eigenstates in quantum Heisenberg models on general graphs.
Our work unifies the approaches of Refs.~\cite{Batista2009,Jepsen2022,Changlani2018,Chertkov2021} by supplying a  graphical interpretation reminiscent of Kirchhoff's law for electrical currents, which we use to solve the generally appearing coupled trigonometric equations \cite{Cerezo2015,Cerezo2016,Cerezo2017}.
According to these rules, solving for product eigenstates amounts to choosing consistent flow patterns of spin supercurrent in the graph.
In particular, we construct Heisenberg models with large degenerate eigenspaces containing a considerable number of product eigenstates. 
To this end, our work builds on Cerezo et al.~\cite{Cerezo2015}, who classified product eigenstates in spin-$1/2$ systems with non-uniform exchange interactions and magnetic fields in order to systematically engineer models with fully factorized ground states.
We find that for models such as the square lattice with zig-zag edge the degeneracy scales exponentially in the size of the boundary of the system, and
describe a suspension procedure to construct spin systems with a degeneracy of the product eigenstate subspace growing exponentially in system size.
Other quantum spin models with extensive degeneracy have been recently proposed \cite{Changlani2018,Palle2021,Dmitriev2021,Nussinov2023,Fendley2019,Zadnik2021-1,Zadnik2021-2,Pozsgay2021}.

We numerically find that 
the degeneracy of the energy eigenspace that accommodates the product eigenstates exceeds the dimension of the space spanned by the product eigenstates
in accordance with earlier studies \cite{Batista2009,Batista2012,Changlani2018}.
It is built up by additional non-product states that evade a simple description in terms of symmetries, and, in spin ladders, these additional states were linked to an anyonic condensate \cite{Batista2012}.
The class of models with nontrivial product eigenstates that we characterize allow for
significant simplifications
in Bethe ansatz approaches to $XXZ$ quantum spin chains, including the existence of fully factorized pseudo vacua \cite{Cao2003}, homogeneous $T$-$Q$-relations for chains with non-diagonal -- i.e., $U(1)$-symmetry breaking -- boundary conditions \cite{Murgan2005}, and the chiral reformulation of the ordinary Bethe ansatz where the elemental excitations are defect angles in helices \cite{Popkov2021,Zhang2021-1,Zhang2021-2}.
By the general theory and the derived models with extensive degeneracies, spin product eigenstates and Heisenberg models on graphs hint at a crucial role of graph topology for entanglement, phase transitions and exact solvability.
The condensate corresponding to the macroscopically degenerate eigenspace
can be investigated in the described experimental platforms
\cite{Gross2017,Chen2023,Sbierski2023,Ebadi2021,Semeghini2021,Jepsen2022,King2018,Tacchino2020,VanDyke2022,Khajetoorians2011,Khajetoorians2013,Chen2022,Eckardt2010,Struck2011,Struck2013,Vorberg2013,Wu2022,Petiziol2024,Schnell2023}
and could be particularly easily initialized with high fidelity in quantum computers and cold atom systems when compared to entangled states \cite{Jepsen2022}.
After completion of our work, we became aware of the recent Ref.~\cite{Zhang2023}, which touches on similar ideas.

\paragraph{Model \& Method---}

We investigate quantum spins-$1/2$ representation on the vertices of a connected simple undirected graph $\Lambda = (V, E)$, where $V$ is the set of vertices
and $E$ is the set of edges.
This excludes edges connecting the same pair of vertices and loops connecting a vertex to itself.
The dynamics are given in terms of a Hamiltonian constructed by assigning an $XXZ$ spin-exchange interaction term and a Dzyaloshinskii-Moriya interaction (DMI) term to each edge.
\begin{align}
\begin{split}
\label{eq:Hamiltonian}
	H = \sum_{\{ i,j \} \in E} \Big[&J \left( S^x_i S^x_j + S^y_i S^y_j \right) + \Delta S^z_i S^z_j \\
	&- \kappa_{ij} D \left( S^x_i S^y_j - S^y_i S^x_j \right)\Big],
\end{split}
\end{align}
with the exchange interaction $J$, the anisotropy $\Delta$ and the spin-$1/2$ operators $S^{x/y/z}_i = \hbar \sigma_i^{x/y/z} / 2$
acting on the spin at vertex $i$, where $\sigma_i^{x/y/z}$ are Pauli matrices.
$D \geq 0$ is the strength of the DMI and $\kappa_{ij}= - \kappa_{ji} = \pm1$ is its sign.
For $J=D=0$, this is the quantum Ising model where an eigenbasis is given by products of eigenvectors of $\sigma^z$. In the following, we assume $J^2 + D^2 \neq 0$.

The fully polarized states $\ket{\uparrow \dots \uparrow}$ and $\ket{\downarrow \dots \downarrow}$ are product eigenstates  for all Hamiltonians of the form in Eq.~(\ref{eq:Hamiltonian}).
For finding additional product eigenstates, we follow
the approach in \cite{Cerezo2015} and insert the general ansatz for a product state
\begin{equation}
	\label{eq:ProductState}
	\ket{\Psi(\bm{\vartheta}, \bm{\varphi})} = \bigotimes_{i \in V} \left( \cos{\left(\frac{\vartheta_i}{2}\right)} e^{-\mathrm{i} \frac{\varphi_i}{2} } \ket{\uparrow}_i + \sin{\left(\frac{\vartheta_i}{2}\right)} e^{\mathrm{i} \frac{\varphi_i}{2} } \ket{\downarrow}_i\right)
\end{equation}
into the eigenvalue equation for $H$.
Here, $\bm{\vartheta} = (\vartheta_1, \dots, \vartheta_\abs{V})$ are the angles relative to the hard-axis $z$ and $\bm{\varphi} = (\varphi_1, \dots, \varphi_\abs{V})$ are the angles in the easy-plane $x$-$y$, taking values $\vartheta_i \in (0, \pi)$ and $\varphi_i = [0, 2\pi)$.
The eigenvalue problem $H \ket{\Psi(\bm{\vartheta}, \bm{\varphi})} = \varepsilon(\bm{\vartheta}, \bm{\varphi}) \ket{\Psi(\bm{\vartheta}, \bm{\varphi})}$
yields a coupled set of trigonometric equations constraining the angles $\bm{\vartheta}$ and $\bm{\varphi}$, see \cite{Cerezo2015} and the Supplemental Material \footnote{See Supplemental Material at [URL will be inserted by the publisher] for details.}.

\paragraph{Results---}

Among the constraining equations there are pairs, one for each edge, that constrain the relative angles of adjacent spins. The remaining equations can then be condensed to very few rules that characterize all product eigenstates, which we discuss in the following.
For easy-axis magnetism, $\Delta^2 > J^2 + D^2$,
we find that there are only the trivial product eigenstates $\ket{\uparrow \dots \uparrow}$ and $\ket{\downarrow \dots\downarrow}$ \cite{Note1}. Additional interactions such as magnetic fields are required to obtain further product eigenstates in this case.
For easy-plane magnetism, $\Delta^2 \leq J^2 + D^2$, we develop a graphical interpretation of the constraints for the product eigenstates. 
First,
the polar angle of the spins is constant $\vartheta_j = \Theta$, corresponding to a canted spin configuration.
Secondly, the exchange interactions constrain the azimuthal angles of adjacent spins to $\varphi_i-\varphi_j = \kappa_{ij} \delta + \sigma_{ij} \gamma$, where the DMI defines a bias angle
$\delta = \arg\left(J + \mathrm{i} D \right)$
and the anisotropy determines the anisotropy angle $\gamma = \arccos\left(\Delta / \sqrt{J^2 + D^2}\right)$. 
Here, each edge has a binary degree of freedom with $\sigma_{ij} = - \sigma_{ji} = \pm 1$.
As a consequence, the degrees of freedom of the product eigenstates are two continuous global angles $\Theta$ and $\Phi$, fixing the orientation of a reference spin $\vartheta_1 \equiv \Theta$ and $\varphi_1 \equiv \Phi$, as well as the discrete degree of freedom of choosing  signs $\sigma_{ij}$ for the edges. 
The latter degree of freedom corresponds to choosing an orientation of the underlying graph described by assigning a direction to an edge $\{i,j\}$, that we depict by an arrow pointing towards vertex $i$ if 
$\varphi_i - \varphi_j = \kappa_{ij} \delta + \gamma$
and in the other direction if
$\varphi_i - \varphi_j = \kappa_{ij} \delta - \gamma$.
\begin{figure}
	\includegraphics[width=\columnwidth]{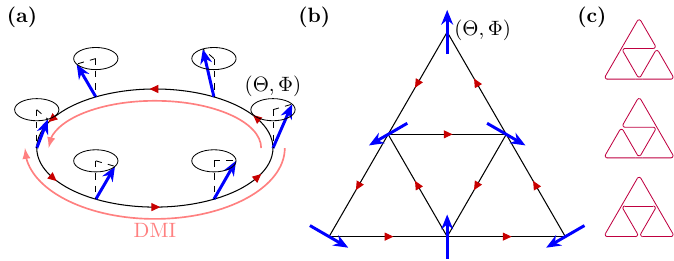}
	\caption{\label{fig:triangle} Graphical solution to the product eigenstate problem, shown by a Heisenberg model on two graphs with spins at the vertices (blue arrows depict spin expectation values $\langle \mathbf{S} \rangle = \left(\langle S^x\rangle, \langle S^y\rangle, \langle S^z\rangle\right)^T$) and edges denote the exchange interaction and the DMI, see Eq.(\ref{eq:Hamiltonian}).
	The change of the azimuthal angle along an edge can take the values $\kappa_{ij} \delta + \sigma_{ij} \gamma$. 	The sign $\sigma_{ij}$ is indicated by the direction of the red arrows. 
	\textbf{(a)} Spin chain with anisotropy and bias angles $\gamma = \pi /3$ and $\delta = \pi /6$.
	The signs of the DMI $\kappa_{ij}$ are indicated by rose arrows giving bias angles of $+\delta$ (top half) and $-\delta$ (bottom half).
	The reference angles are $\Theta=\pi/3$ and $\Phi=\pi/6$.
	\textbf{(b)} Triangular lattice with $\gamma = 2\pi /3$ and $D=0$ with a product eigenstate 
	and the corresponding
	Kirchhoff orientation
	(red arrows)
	for reference angles $\Theta=\Phi=\pi/2$ (top view).
	\textbf{(c)} Euler circuit for the triangular lattice in (b).
	}
\end{figure}
The choice of the orientation $\sigma$ has to be commensurate with two rules reminiscent of Kirchhoff's laws for electrical currents:
\begin{itemize}[itemindent=4em]
	\item[
	\textbf{Vertex rule}:
	]
	At each vertex $i$, the number of ingoing arrows equals the number of outgoing arrows. $\sum_{\{i,j\}\in E} \sigma_{ij}
	= 0$, depicted by, e.g.,  
	$
    \begin{array}{l}
    \includegraphics[width=40pt]{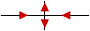}
    \end{array}
    $.
	\item[
	\textbf{Circuit rule}:
	]
	The change of the azimuthal angle along a circuit $\Gamma \subset E$ in the graph must consistently be an integer multiple of $2\pi$. $\sum_{\{i,j\}\in \Gamma} \left(\kappa_{ij} \delta + \sigma_{ij} \gamma\right)
	\equiv 0 \ (\operatorname{mod}  2\pi)$, depicted by, e.g.,  
$
\begin{array}{l}
\includegraphics[width=40pt]{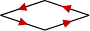}
\end{array}
$.
\end{itemize}
There are two special cases for $\Delta^2 = J^2 + D^2$.
First, for $\gamma=0$, $\sigma$ is not well-defined since $-\gamma \equiv \gamma \ (\operatorname{mod} 2\pi)$.
Yet, the circuit rule remains meaningful and allows for nontrivial product eigenstates only at special values of $\delta$.
By means of local rotations about the anisotropy axis, these Hamiltonians are unitarily equivalent to the XXX Hamiltonian \cite{Perk1976} whose product eigenstates are fully polarized states along an arbitrary direction.
Second, for $\gamma=\pi$, $\sigma$ is not well-defined either.
Still, like in the previous case, nontrivial product eigenstates exist if the circuit rule can be fulfilled.
The vertex rule is not to be confused with Gauss' law in lattice gauge theories \cite{Chandrasekharan1997}.
In physical terms, the vertex rule corresponds to the local conservation of the $z$-component of the spin, i.e., compensating spin supercurrents $C_{ij}$ at vertex $i$
\begin{equation}
\label{eq:SpinSuperCurrent}
	\langle \dot{S^z_i} \rangle
	=  \hspace*{-1ex} \sum_{\{i, j\} \in E} \hspace*{-1ex} C_{ij} =\frac{\sin^2(\Theta)\sin(\gamma)\hbar^2 J}{4\cos(\delta)} \hspace*{-1ex} \sum_{\{i, j\} \in E} \hspace*{-1ex} \sigma_{ij}
	 = 0.
\end{equation}
In accordance with this interpretation,
the vertex rule can be extended to Hamiltonians including local magnetic fields $\mathbf{B}_i \mathbf{S}_i$ such that the fields act as source and sink terms compensating the difference between ingoing and outgoing arrows \cite{Note1}.
Fine-tuning the DMI \cite{Kim2013,Shanavas2014,Zhang2018} and local magnetic fields can hence help to fulfill the circuit and vertex rule, respectively. 

An orientation $\sigma$ that satisfies the vertex rule at each vertex is called an \textit{Euler orientation}.
Graphs with Euler orientations have at least one closed path that travels each edge in the direction of the Euler orientation exactly once. Such a path can be explicitly constructed, e.g., by the Hierholzer algorithm \cite{Euler1741,Gibbons1985}.
To simplify terminology, we refer to an Euler orientation fulfilling the circuit rule for a Hamiltonian in Eq.~\eqref{eq:Hamiltonian} as a \textit{Kirchhoff orientation} of a \textit{Kirchhoff graph}.
This should not be confused with the notion of Kirchhoff graphs in the context of reaction route graphs~\cite{Fehribach2009,Fishtik2004}.
Any product state commensurate with the Kirchhoff rules has an appealing interpretation: When we unravel a closed path as above,
the state looks like a helix with local pitch angles $\gamma$ 
that are distorted by local bias angles $\pm\delta$ in presence of DMI.
In this sense, the product states are a natural extension of the recently discovered phantom helices~\cite{Popkov2021,Zhang2021-1,Zhang2021-2} to arbitrary graphs and dimensions.
Remarkably, all product eigenstates of Eq.~\eqref{eq:Hamiltonian}
have the same energy independent of DMI 
\begin{equation}
\label{eq:Energy}
	\varepsilon 
	= \hbar^2 \Delta \abs{E} /4,
\end{equation}
with $\abs{E}$ the number of edges.
We note that these results carry over to higher spin representations when replacing the ansatz in Eq.~\eqref{eq:ProductState} by a product of spin-coherent states. These states are still parameterized by two angles obeying the Kirchhoff laws and populating an energetically degenerate subspace.
In the following, we refer to the space of all eigenstates with the same energy as the product eigenstates as the \textit{$\varepsilon$-space} and the subspace spanned by the product eigenstates as \textit{product eigenspace}.
An upper bound for the product eigenspace dimension 
(spin-$1/2$)
is given by
\begin{equation}
    d_\text{max} = 2 +
    \sum_{k=1}^{\abs{V}-1} \operatorname{min}\left(\abs{\sigma}, \begin{pmatrix} \abs{V} \\ k\end{pmatrix}\right),
\end{equation}
where $\abs{\sigma}$ is the number of Kirchhoff orientations and $\abs{V}$ the number of spins.
For $\gamma=0$ and $\gamma=\pi$, we set $\abs{\sigma}=1$.
This bound follows from considering the projections on the $\abs{V}+1$ eigenspaces $P_k$, $k\in\{0, ... \abs{V}\}$, of the total $z$-spin symmetry operator $\sum_{i \in V} S^z_i$.
By Eq.~\eqref{eq:ProductState}, a product eigenstate $\ket{\Psi(\Theta, \Phi)}$ with $\Theta\neq 0, \pi$ gives $\abs{V}+1$ orthogonal eigenstates $P_k \ket{\Psi(\Theta, \Phi)}$ of the same energy.
Given another nontrivial product eigenstate $\ket{\Psi(\Theta^\prime, \Phi^\prime)}$ with the same Kirchhoff orientation, the projected states differ merely by a non-zero factor, i.e., $P_k \ket{\Psi(\Theta^\prime, \Phi^\prime)} \sim P_k \ket{\Psi(\Theta, \Phi)}$.
So, for any projector $P_k$, the image of the product eigenspace under this projector contains at most $\abs{\sigma}$ linear independent vectors and is further bounded by the dimension of $\operatorname{im}(P_k)$ which is $\abs{V}$ choose $k$.
Since the trivial product eigenstates $\ket{\uparrow \dots \uparrow}$ and $\ket{\downarrow \dots \downarrow}$ span the one-dimensional spaces $\operatorname{im}(P_0)$ and $\operatorname{im}(P_\abs{V})$, $d_\text{max}$ is at least $2$.
A general lower bound of $\abs{V} + 1$ originates from the $U(1)$-symmetry if at least one nontrivial product eigenstate exists since such a state has non-zero weight in every space $\operatorname{im}(P_k)$.
For graphs with vanishing DMI and finite anisotropy angle $\gamma = \pi / (2n)$, $n$ a positive integer, the number of Kirchhoff orientations $\abs{\sigma}$ gives an improved lower bound on the product eigenspace degeneracy.
In this case, the dimension of the product eigenspace has both an upper and a lower bound linear in $\abs{\sigma}$.
For details, see the Supplemental Material \cite{Note1}.

We can interpret the product eigenstates physically in terms of delocalized spin-flips.
For a given Kirchhoff orientation, the span of the product eigenstates $\ket{\Psi(\Theta, \Phi)}$
is generated by states $R^{(n)}\ket{\Psi(\pi/2, \Phi_0)}$ consisting of $n$ delocalized spin-flips with $0 \leq n \leq \abs{V}$ and a fixed $\Phi_0$.
Here,
\begin{equation}
    R^{(n)} = \sum_{\{i_1, \dots i_n \} \subset V } P_{i_1} \dots P_{i_n}
\end{equation}
where $P_i$ acts on a product eigenstate as in Eq.~\eqref{eq:ProductState} by $\vartheta_i \mapsto \pi - \vartheta_i$ and $\varphi_i \mapsto \varphi_i - \pi$ and hence flips the expectation value of the $i^\text{th}$ spin, i.e., $\langle \mathbf{S}_i \rangle \mapsto -\langle \mathbf{S}_i \rangle$.
We set $R^{(0)} \equiv 1$ and note that $R^{(n)} = 0$ for $n>\abs{V}$ since the sum is empty.
The spin-flip states can be alternatively expressed in terms of a linear combination of derivatives of $\ket{\Psi(\Theta, \Phi)}$ with respect to $\Theta$ and $\Phi$
which are also eigenstates of $H$, given by Eq.~\eqref{eq:Energy},
\begin{align}
\label{eq:theta_der}
	\partial_\Theta \ket{\Psi(\Theta, \Phi)} &= \frac{\mathrm{i}}{2} \sum_{i \in V} P_i \ket{\Psi(\Theta, \Phi)}, \\
\label{eq:phi_der}
	\partial_\Phi \ket{\Psi(\Theta, \Phi)} &= \frac{1}{2} \sum_{i \in V} 
	e^{-\frac{\mathrm{i}}{\hbar} \pi S_i^z} \ket{\Psi(\Theta, \Phi)},
\end{align}
and at $\Theta=\pi/2$, $\Phi=\Phi_0$ the derivatives can be conversely expressed in terms of the spin-flip states. 
The delocalized spin-flip states then span the product eigenspace because
the eigenspace generated by all product eigenstates
and the eigenspace generated by the derivatives at the angles $\Theta=\pi/2$ and $\Phi=\Phi_0$ are the same. 
This is because on the one hand every product state has a converging Taylor expansion in $\Theta$ and $\Phi$ 
(see Eq.~\eqref{eq:ProductState} with $\vartheta_1 = \Theta$, $\varphi_1 = \Phi$ and the other angles according to the Kirchhoff orientation),
and on the other hand, every derivative can be expressed as the limit of a differential quotient of product eigenstates.

\paragraph{Applications---}

\begin{figure*}
	\includegraphics[width=\linewidth]{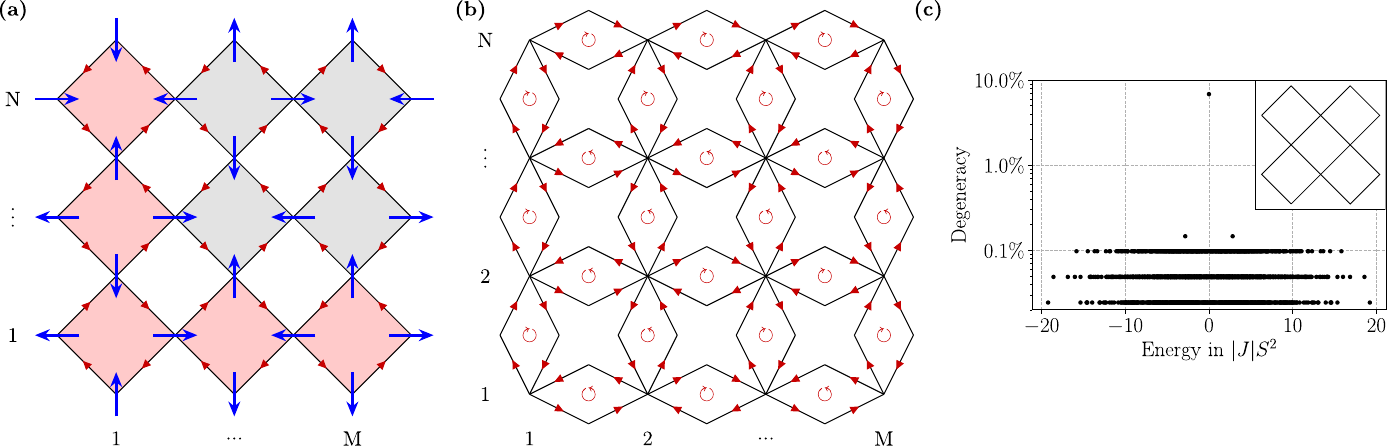}
	\caption{\label{fig:jaeger}
	\textbf{(a)} $XX$ model ($\Delta = D = 0$) on a square lattice with zig-zag edge and its product eigenstates.
	Fixing $\Theta=\pi/2$, the spins align row- and column-wise in a Néel pattern determined by the boundary spins for the $N$ rows and $M$ columns resulting in $2^{N+M-1}$ Kirchhoff orientations.
	\textbf{(b)} 
	The replacement of edges by ``diamond" squares allows for a number of Kirchhoff orientations that grows exponentially in the number of diamonds with a lower bound of $2^{2NM+N+M}$ Kirchhoff orientations.
	\textbf{(c)} Degeneracy of the eigenspaces of the $XX$ model on a square lattice, see inset,
	as a fraction of the total Hilbert space dimension.
	The eigenspace that accommodates the product eigenstates has energy zero commensurate with Eq.~\eqref{eq:Energy}, and is two orders of magnitude more degenerate than typical eigenspaces.
    ($S=\hbar /2$)
    }
\end{figure*}


We next discuss well-known systems without DMI and their product eigenstates in our graphical framework.
In 1D, spin helices are well-studied product eigenstates.
For vanishing DMI, a periodic spin chain with $N$ sites has two Kirchhoff orientations if the periodic closure condition $\gamma N \equiv 0 \ (\operatorname{mod} 2\pi)$ \cite{Cerezo2016,Popkov2021} is fulfilled,
corresponding to two helices with opposite helicity.
In absence of DMI, inverting the direction of arrows in a Kirchhoff orientation generally yields another permissible orientation.
Numerically, we find that the product eigenstates in the circular chain generate a degeneracy linear in the chain length $N$
\cite{Note1}.
However, the product eigenstates do not span the entire $\varepsilon$-space for all choices of permissible $\gamma$ and $N$.
The triangular lattice depicted in Fig.~\ref{fig:triangle}b also allows for two different choices of Kirchhoff orientations independent of the system size whose product eigenstates are part of the ground space \cite{Changlani2018}. More generally for graphs made up of triangular motifs, the choice of a Kirchhoff orientation is equivalent to a three-coloring of the graph which has been used to study product eigenstates in such systems \cite{Changlani2018}.
For the kagome lattice, the abundance of three-colorings yields an extensively degenerate ground space, i.e., for this family of systems, the degeneracy grows exponentially in the number of spins.
We proceed with constructing novel lattices with macroscopic degeneracy.
We first detail the square lattice with zig-zag boundary conditions, see Fig.~\ref{fig:jaeger}a, where the number of Kirchhoff orientations increases exponentially with the size of the system's boundary.
To that end, we consider the $XX$ model, i.e., $\Delta = D = 0$ in Eq.~\eqref{eq:Hamiltonian}, corresponding to $\gamma = \pi/2$. 
We find that due to the zig-zag boundary, the permissible Kirchhoff orientations form checkerboard patterns of vortices, c.f.\ Fig.~\ref{fig:jaeger}a where arrows go either clockwise or counter-clockwise around the colored plaquettes.
Specifying the $M$ vortices at the bottom and the $N-1$ vortices at the left boundary  (rose in Fig.~\ref{fig:jaeger}a) fully determines the inner ones (gray and white).
Hence, the number of Kirchhoff orientations is $2^{N+M-1}$, i.e., grows exponentially with the size of the boundary \cite{Note1}.
On the level of the individual spins,
fixing the spin in the lower left corner to point along the $x$-axis,
the degrees of freedom are choosing the leftmost spins pointing either in the $(+x)$- or $(-x)$-direction and the bottom spins pointing either in the $(+y)$- or $(-y)$-direction.
The remaining spins anti-align with their neighbors along each column and row of spins, see Fig.~\ref{fig:jaeger}a.
We find numerically that the dimension of the space spanned by product eigenstates is consistent with $(M N + 1) 2^{N + M}$.
In the special case of $N=1$ or $M=1$, the degeneracy scales exponentially with the system's size, and is hence an extensive quantity.
Such diamond chains have been studied among other reasons for their degeneracy in the context of exactly solvable systems \cite{Sutherland1983,Takano1996} and spin ladders \cite{Ishii2000}. 
We find that systems with an extensively degenerate eigenspace of product eigenstates 
can be engineered by a suspension procedure similar to earlier ideas of building graphs from smaller motifs \cite{Changlani2018,Chertkov2021,Strecka2002,Caci2023}.
For a given graph,
we replace each edge of the graph by a subgraph
in a way that respects the Kirchhoff rules, e.g., for graphs with all circuits of even length like the square lattice, we substitute the edges by a single diamond square oriented clockwise or anticlockwise, see Fig.~\ref{fig:jaeger}b.
The construction respects the vertex and the circuit rule for $\gamma=\pi/2$ and 
yields extensive degeneracy, exponential in the number of independently oriented diamonds. This results in a lower bound of $2^{2NM+N+M}$ in the case of the square lattice with $N$ and $M$ the number of rows and columns of the square lattice before the suspension, respectively.
A variety of systems with an extensive number of Kirchhoff orientations can be constructed by removing arbitrary vertices and corresponding edges from the square lattice before suspending the diamonds.

\paragraph{Degeneracy \& intrinsic structure of $\varepsilon$-spaces---}
In general, the degeneracy of the product eigenspace exceeds the number of possible Kirchhoff orientations.
We analyze the $\varepsilon$-space for the square lattice and the diamond chain. 
We numerically confirm for $(N, M) = (1, 1), (1,2), (1, 3)$, and $(2, 2)$ that the $\varepsilon$-space is the most degenerate eigenspace \cite{Note1}.
Typically the eigenspaces are one or two orders of magnitude more degenerate than all other eigenspaces, making up a large share of the total Hilbert space dimension, see 
Fig.~\ref{fig:jaeger}c. 
Additional degeneracy stems from states with non-vanishing entanglement entropy.
In special cases in 1D, these have been associated with anyonic condensates \cite{Batista2009} or spin liquid phases \cite{Changlani2018}.

\paragraph{Discussion---}

We developed the complete classification scheme of product eigenstates of Heisenberg Hamiltonians with $XXZ$ exchange interactions and DMI by Kirchhoff rules for the underlying graph.
We thereby give a concrete example for a fundamental property connecting graph topology, degeneracy, and entanglement.
Some graphs feature associated Kirchhoff orientations in abundance.
They possess large degenerate eigenspaces with zero-energy delocalized spin-flip excitations and additional non-product states associated to anyonic statistics in special systems.
The degeneracy is reduced when moving away from the critical value of $\gamma$ \cite{Note1}.
The singular nature of the degeneracy of the eigenspace accommodating the product eigenstates is therefore not explained by the natural symmetries of the model.
The $\varepsilon$-space that appears as a condensate may be especially accessible by optically driving the system.
There, the extensive degeneracy compensates the reduced transition rates caused by the energy difference which results in a macroscopic observable.
In conjunction with a recent prediction of one-dimensional anyon condensates \cite{Batista2012}, our work suggests exotic characteristics of the related condensates in higher-dimensional systems, and experimental investigations could unveil exotic quasiparticles which are not accessible by analytical techniques for higher-dimensional systems.
The considered graphs in general describe quantum spin systems of any dimension.
As such, our study could be a step towards a partial solution of $XXZ$ Heisenberg models in more than one dimension for a critical
value of $\gamma$, since we have shown that product eigenstates can be interpreted as spin helices along a one-dimensional Euler circuit. At special anisotropies these paths may allow 
the application of ideas from well-established one-dimensional techniques like Jordan-Wigner transforms \cite{Jordan1928,Lieb1961,Fradkin1989} and bosonization approaches \cite{Delft1998} to higher-dimensional models.

\begin{acknowledgments}
FG\ and TP\ thank Oleg Tchernyshyov, and TP\ thanks Mikhail Ivanov for helpful discussions.
FG\ acknowledges funding by the Cluster of Excellence ``CUI: Advanced Imaging of Matter'' of the Deutsche Forschungsgemeinschaft (DFG) – EXC 2056 – project ID 390715994.
IR\ and CS\ are partially supported by the Deutsche Forschungsgemeinschaft via the Cluster of Excellence EXC 2121 ``Quantum Universe'' - 390833306.
CS\ is supported by the Deutsche Forschungsgemeinschaft (DFG, German Research Foundation) under SCHW1162/6-1.
TP\ acknowledges funding by the DFG (German Science Foundation) project number 420120155 and the European Union (ERC, QUANTWIST, project number 101039098). Views and opinions expressed are however those of the authors only and do not necessarily reflect those of the European Union or the European Research Council. Neither the European Union nor the granting authority can be held responsible for them.
\end{acknowledgments}



\begin{thebibliography}{77}%
\makeatletter
\providecommand \@ifxundefined [1]{%
 \@ifx{#1\undefined}
}%
\providecommand \@ifnum [1]{%
 \ifnum #1\expandafter \@firstoftwo
 \else \expandafter \@secondoftwo
 \fi
}%
\providecommand \@ifx [1]{%
 \ifx #1\expandafter \@firstoftwo
 \else \expandafter \@secondoftwo
 \fi
}%
\providecommand \natexlab [1]{#1}%
\providecommand \enquote  [1]{``#1''}%
\providecommand \bibnamefont  [1]{#1}%
\providecommand \bibfnamefont [1]{#1}%
\providecommand \citenamefont [1]{#1}%
\providecommand \href@noop [0]{\@secondoftwo}%
\providecommand \href [0]{\begingroup \@sanitize@url \@href}%
\providecommand \@href[1]{\@@startlink{#1}\@@href}%
\providecommand \@@href[1]{\endgroup#1\@@endlink}%
\providecommand \@sanitize@url [0]{\catcode `\\12\catcode `\$12\catcode
  `\&12\catcode `\#12\catcode `\^12\catcode `\_12\catcode `\%12\relax}%
\providecommand \@@startlink[1]{}%
\providecommand \@@endlink[0]{}%
\providecommand \url  [0]{\begingroup\@sanitize@url \@url }%
\providecommand \@url [1]{\endgroup\@href {#1}{\urlprefix }}%
\providecommand \urlprefix  [0]{URL }%
\providecommand \Eprint [0]{\href }%
\providecommand \doibase [0]{https://doi.org/}%
\providecommand \selectlanguage [0]{\@gobble}%
\providecommand \bibinfo  [0]{\@secondoftwo}%
\providecommand \bibfield  [0]{\@secondoftwo}%
\providecommand \translation [1]{[#1]}%
\providecommand \BibitemOpen [0]{}%
\providecommand \bibitemStop [0]{}%
\providecommand \bibitemNoStop [0]{.\EOS\space}%
\providecommand \EOS [0]{\spacefactor3000\relax}%
\providecommand \BibitemShut  [1]{\csname bibitem#1\endcsname}%
\let\auto@bib@innerbib\@empty
\bibitem [{\citenamefont {Gross}\ and\ \citenamefont
  {Bloch}(2017)}]{Gross2017}%
  \BibitemOpen
  \bibfield  {author} {\bibinfo {author} {\bibfnamefont {C.}~\bibnamefont
  {Gross}}\ and\ \bibinfo {author} {\bibfnamefont {I.}~\bibnamefont {Bloch}},\
  }\bibfield  {title} {\bibinfo {title} {Quantum simulations with ultracold
  atoms in optical lattices},\ }\href@noop {} {\bibfield  {journal} {\bibinfo
  {journal} {Science}\ }\textbf {\bibinfo {volume} {357}},\ \bibinfo {pages}
  {995} (\bibinfo {year} {2017})}\BibitemShut {NoStop}%
\bibitem [{\citenamefont {Chen}\ \emph
  {et~al.}(2023{\natexlab{a}})\citenamefont {Chen}, \citenamefont {Bornet},
  \citenamefont {Bintz}, \citenamefont {Emperauger}, \citenamefont {Leclerc},
  \citenamefont {Liu}, \citenamefont {Scholl}, \citenamefont {Barredo},
  \citenamefont {Hauschild}, \citenamefont {Chatterjee}, \citenamefont
  {Schuler}, \citenamefont {L{\"a}uchli}, \citenamefont {Zaletel},
  \citenamefont {Lahaye}, \citenamefont {Yao},\ and\ \citenamefont
  {Browaeys}}]{Chen2023}%
  \BibitemOpen
  \bibfield  {author} {\bibinfo {author} {\bibfnamefont {C.}~\bibnamefont
  {Chen}}, \bibinfo {author} {\bibfnamefont {G.}~\bibnamefont {Bornet}},
  \bibinfo {author} {\bibfnamefont {M.}~\bibnamefont {Bintz}}, \bibinfo
  {author} {\bibfnamefont {G.}~\bibnamefont {Emperauger}}, \bibinfo {author}
  {\bibfnamefont {L.}~\bibnamefont {Leclerc}}, \bibinfo {author} {\bibfnamefont
  {V.~S.}\ \bibnamefont {Liu}}, \bibinfo {author} {\bibfnamefont
  {P.}~\bibnamefont {Scholl}}, \bibinfo {author} {\bibfnamefont
  {D.}~\bibnamefont {Barredo}}, \bibinfo {author} {\bibfnamefont
  {J.}~\bibnamefont {Hauschild}}, \bibinfo {author} {\bibfnamefont
  {S.}~\bibnamefont {Chatterjee}}, \bibinfo {author} {\bibfnamefont
  {M.}~\bibnamefont {Schuler}}, \bibinfo {author} {\bibfnamefont {A.~M.}\
  \bibnamefont {L{\"a}uchli}}, \bibinfo {author} {\bibfnamefont {M.~P.}\
  \bibnamefont {Zaletel}}, \bibinfo {author} {\bibfnamefont {T.}~\bibnamefont
  {Lahaye}}, \bibinfo {author} {\bibfnamefont {N.~Y.}\ \bibnamefont {Yao}},\
  and\ \bibinfo {author} {\bibfnamefont {A.}~\bibnamefont {Browaeys}},\
  }\bibfield  {title} {\bibinfo {title} {Continuous symmetry breaking in a
  two-dimensional {R}ydberg array},\ }\href
  {https://doi.org/10.1038/s41586-023-05859-2} {\bibfield  {journal} {\bibinfo
  {journal} {Nature}\ }\textbf {\bibinfo {volume} {616}},\ \bibinfo {pages}
  {691} (\bibinfo {year} {2023}{\natexlab{a}})}\BibitemShut {NoStop}%
\bibitem [{\citenamefont {Sbierski}\ \emph {et~al.}(2024)\citenamefont
  {Sbierski}, \citenamefont {Bintz}, \citenamefont {Chatterjee}, \citenamefont
  {Schuler}, \citenamefont {Yao},\ and\ \citenamefont {Pollet}}]{Sbierski2023}%
  \BibitemOpen
  \bibfield  {author} {\bibinfo {author} {\bibfnamefont {B.}~\bibnamefont
  {Sbierski}}, \bibinfo {author} {\bibfnamefont {M.}~\bibnamefont {Bintz}},
  \bibinfo {author} {\bibfnamefont {S.}~\bibnamefont {Chatterjee}}, \bibinfo
  {author} {\bibfnamefont {M.}~\bibnamefont {Schuler}}, \bibinfo {author}
  {\bibfnamefont {N.~Y.}\ \bibnamefont {Yao}},\ and\ \bibinfo {author}
  {\bibfnamefont {L.}~\bibnamefont {Pollet}},\ }\bibfield  {title} {\bibinfo
  {title} {Magnetism in the two-dimensional dipolar {XY} model},\ }\href
  {https://doi.org/10.1103/PhysRevB.109.144411} {\bibfield  {journal} {\bibinfo
   {journal} {Phys. Rev. B}\ }\textbf {\bibinfo {volume} {109}},\ \bibinfo
  {pages} {144411} (\bibinfo {year} {2024})}\BibitemShut {NoStop}%
\bibitem [{\citenamefont {Ebadi}\ \emph {et~al.}(2021)\citenamefont {Ebadi},
  \citenamefont {Wang}, \citenamefont {Levine}, \citenamefont {Keesling},
  \citenamefont {Semeghini}, \citenamefont {Omran}, \citenamefont {Bluvstein},
  \citenamefont {Samajdar}, \citenamefont {Pichler}, \citenamefont {Ho},
  \citenamefont {Choi}, \citenamefont {Sachdev}, \citenamefont {Greiner},
  \citenamefont {Vuletić},\ and\ \citenamefont {Lukin}}]{Ebadi2021}%
  \BibitemOpen
  \bibfield  {author} {\bibinfo {author} {\bibfnamefont {S.}~\bibnamefont
  {Ebadi}}, \bibinfo {author} {\bibfnamefont {T.~T.}\ \bibnamefont {Wang}},
  \bibinfo {author} {\bibfnamefont {H.}~\bibnamefont {Levine}}, \bibinfo
  {author} {\bibfnamefont {A.}~\bibnamefont {Keesling}}, \bibinfo {author}
  {\bibfnamefont {G.}~\bibnamefont {Semeghini}}, \bibinfo {author}
  {\bibfnamefont {A.}~\bibnamefont {Omran}}, \bibinfo {author} {\bibfnamefont
  {D.}~\bibnamefont {Bluvstein}}, \bibinfo {author} {\bibfnamefont
  {R.}~\bibnamefont {Samajdar}}, \bibinfo {author} {\bibfnamefont
  {H.}~\bibnamefont {Pichler}}, \bibinfo {author} {\bibfnamefont {W.~W.}\
  \bibnamefont {Ho}}, \bibinfo {author} {\bibfnamefont {S.}~\bibnamefont
  {Choi}}, \bibinfo {author} {\bibfnamefont {S.}~\bibnamefont {Sachdev}},
  \bibinfo {author} {\bibfnamefont {M.}~\bibnamefont {Greiner}}, \bibinfo
  {author} {\bibfnamefont {V.}~\bibnamefont {Vuletić}},\ and\ \bibinfo
  {author} {\bibfnamefont {M.~D.}\ \bibnamefont {Lukin}},\ }\bibfield  {title}
  {\bibinfo {title} {Quantum phases of matter on a 256-atom programmable
  quantum simulator},\ }\href@noop {} {\bibfield  {journal} {\bibinfo
  {journal} {Nature}\ }\textbf {\bibinfo {volume} {595}},\ \bibinfo {pages}
  {227} (\bibinfo {year} {2021})}\BibitemShut {NoStop}%
\bibitem [{\citenamefont {Semeghini}\ \emph {et~al.}(2021)\citenamefont
  {Semeghini}, \citenamefont {Levine}, \citenamefont {Keesling}, \citenamefont
  {Ebadi}, \citenamefont {Wang}, \citenamefont {Bluvstein}, \citenamefont
  {Verresen}, \citenamefont {Pichler}, \citenamefont {Kalinowski},
  \citenamefont {Samajdar}, \citenamefont {Omran}, \citenamefont {Sachdev},
  \citenamefont {Vishwanath}, \citenamefont {Greiner}, \citenamefont
  {Vuletić},\ and\ \citenamefont {Lukin}}]{Semeghini2021}%
  \BibitemOpen
  \bibfield  {author} {\bibinfo {author} {\bibfnamefont {G.}~\bibnamefont
  {Semeghini}}, \bibinfo {author} {\bibfnamefont {H.}~\bibnamefont {Levine}},
  \bibinfo {author} {\bibfnamefont {A.}~\bibnamefont {Keesling}}, \bibinfo
  {author} {\bibfnamefont {S.}~\bibnamefont {Ebadi}}, \bibinfo {author}
  {\bibfnamefont {T.~T.}\ \bibnamefont {Wang}}, \bibinfo {author}
  {\bibfnamefont {D.}~\bibnamefont {Bluvstein}}, \bibinfo {author}
  {\bibfnamefont {R.}~\bibnamefont {Verresen}}, \bibinfo {author}
  {\bibfnamefont {H.}~\bibnamefont {Pichler}}, \bibinfo {author} {\bibfnamefont
  {M.}~\bibnamefont {Kalinowski}}, \bibinfo {author} {\bibfnamefont
  {R.}~\bibnamefont {Samajdar}}, \bibinfo {author} {\bibfnamefont
  {A.}~\bibnamefont {Omran}}, \bibinfo {author} {\bibfnamefont
  {S.}~\bibnamefont {Sachdev}}, \bibinfo {author} {\bibfnamefont
  {A.}~\bibnamefont {Vishwanath}}, \bibinfo {author} {\bibfnamefont
  {M.}~\bibnamefont {Greiner}}, \bibinfo {author} {\bibfnamefont
  {V.}~\bibnamefont {Vuletić}},\ and\ \bibinfo {author} {\bibfnamefont
  {M.~D.}\ \bibnamefont {Lukin}},\ }\bibfield  {title} {\bibinfo {title}
  {Probing topological spin liquids on a programmable quantum simulator},\
  }\href {https://doi.org/10.1126/science.abi8794} {\bibfield  {journal}
  {\bibinfo  {journal} {Science}\ }\textbf {\bibinfo {volume} {374}},\ \bibinfo
  {pages} {1242} (\bibinfo {year} {2021})}\BibitemShut {NoStop}%
\bibitem [{\citenamefont {Jepsen}\ \emph {et~al.}(2022)\citenamefont {Jepsen},
  \citenamefont {Lee}, \citenamefont {Lin}, \citenamefont {Dimitrova},
  \citenamefont {Margalit}, \citenamefont {Ho},\ and\ \citenamefont
  {Ketterle}}]{Jepsen2022}%
  \BibitemOpen
  \bibfield  {author} {\bibinfo {author} {\bibfnamefont {P.~N.}\ \bibnamefont
  {Jepsen}}, \bibinfo {author} {\bibfnamefont {Y.~K.~E.}\ \bibnamefont {Lee}},
  \bibinfo {author} {\bibfnamefont {H.}~\bibnamefont {Lin}}, \bibinfo {author}
  {\bibfnamefont {I.}~\bibnamefont {Dimitrova}}, \bibinfo {author}
  {\bibfnamefont {Y.}~\bibnamefont {Margalit}}, \bibinfo {author}
  {\bibfnamefont {W.~W.}\ \bibnamefont {Ho}},\ and\ \bibinfo {author}
  {\bibfnamefont {W.}~\bibnamefont {Ketterle}},\ }\bibfield  {title} {\bibinfo
  {title} {Long-lived phantom helix states in {H}eisenberg quantum magnets},\
  }\href {https://doi.org/10.1038/s41567-022-01651-7} {\bibfield  {journal}
  {\bibinfo  {journal} {Nat. Phys.}\ }\textbf {\bibinfo {volume} {18}},\
  \bibinfo {pages} {899} (\bibinfo {year} {2022})}\BibitemShut {NoStop}%
\bibitem [{\citenamefont {King}\ \emph {et~al.}(2018)\citenamefont {King},
  \citenamefont {Carrasquilla}, \citenamefont {Raymond}, \citenamefont
  {Ozfidan}, \citenamefont {Andriyash}, \citenamefont {Berkley}, \citenamefont
  {Reis}, \citenamefont {Lanting}, \citenamefont {Harris}, \citenamefont
  {Altomare}, \citenamefont {Boothby}, \citenamefont {Bunyk}, \citenamefont
  {Enderud}, \citenamefont {Fr{\'e}chette}, \citenamefont {Hoskinson},
  \citenamefont {Ladizinsky}, \citenamefont {Oh}, \citenamefont
  {Poulin-Lamarre}, \citenamefont {Rich}, \citenamefont {Sato}, \citenamefont
  {Smirnov}, \citenamefont {Swenson}, \citenamefont {Volkmann}, \citenamefont
  {Whittaker}, \citenamefont {Yao}, \citenamefont {Ladizinsky}, \citenamefont
  {Johnson}, \citenamefont {Hilton},\ and\ \citenamefont {Amin}}]{King2018}%
  \BibitemOpen
  \bibfield  {author} {\bibinfo {author} {\bibfnamefont {A.~D.}\ \bibnamefont
  {King}}, \bibinfo {author} {\bibfnamefont {J.}~\bibnamefont {Carrasquilla}},
  \bibinfo {author} {\bibfnamefont {J.}~\bibnamefont {Raymond}}, \bibinfo
  {author} {\bibfnamefont {I.}~\bibnamefont {Ozfidan}}, \bibinfo {author}
  {\bibfnamefont {E.}~\bibnamefont {Andriyash}}, \bibinfo {author}
  {\bibfnamefont {A.}~\bibnamefont {Berkley}}, \bibinfo {author} {\bibfnamefont
  {M.}~\bibnamefont {Reis}}, \bibinfo {author} {\bibfnamefont {T.}~\bibnamefont
  {Lanting}}, \bibinfo {author} {\bibfnamefont {R.}~\bibnamefont {Harris}},
  \bibinfo {author} {\bibfnamefont {F.}~\bibnamefont {Altomare}}, \bibinfo
  {author} {\bibfnamefont {K.}~\bibnamefont {Boothby}}, \bibinfo {author}
  {\bibfnamefont {P.~I.}\ \bibnamefont {Bunyk}}, \bibinfo {author}
  {\bibfnamefont {C.}~\bibnamefont {Enderud}}, \bibinfo {author} {\bibfnamefont
  {A.}~\bibnamefont {Fr{\'e}chette}}, \bibinfo {author} {\bibfnamefont
  {E.}~\bibnamefont {Hoskinson}}, \bibinfo {author} {\bibfnamefont
  {N.}~\bibnamefont {Ladizinsky}}, \bibinfo {author} {\bibfnamefont
  {T.}~\bibnamefont {Oh}}, \bibinfo {author} {\bibfnamefont {G.}~\bibnamefont
  {Poulin-Lamarre}}, \bibinfo {author} {\bibfnamefont {C.}~\bibnamefont
  {Rich}}, \bibinfo {author} {\bibfnamefont {Y.}~\bibnamefont {Sato}}, \bibinfo
  {author} {\bibfnamefont {A.~Y.}\ \bibnamefont {Smirnov}}, \bibinfo {author}
  {\bibfnamefont {L.~J.}\ \bibnamefont {Swenson}}, \bibinfo {author}
  {\bibfnamefont {M.~H.}\ \bibnamefont {Volkmann}}, \bibinfo {author}
  {\bibfnamefont {J.}~\bibnamefont {Whittaker}}, \bibinfo {author}
  {\bibfnamefont {J.}~\bibnamefont {Yao}}, \bibinfo {author} {\bibfnamefont
  {E.}~\bibnamefont {Ladizinsky}}, \bibinfo {author} {\bibfnamefont {M.~W.}\
  \bibnamefont {Johnson}}, \bibinfo {author} {\bibfnamefont {J.}~\bibnamefont
  {Hilton}},\ and\ \bibinfo {author} {\bibfnamefont {M.~H.}\ \bibnamefont
  {Amin}},\ }\bibfield  {title} {\bibinfo {title} {Observation of topological
  phenomena in a programmable lattice of 1,800 qubits},\ }\href
  {https://doi.org/10.1038/s41586-018-0410-x} {\bibfield  {journal} {\bibinfo
  {journal} {Nature}\ }\textbf {\bibinfo {volume} {560}},\ \bibinfo {pages}
  {456} (\bibinfo {year} {2018})}\BibitemShut {NoStop}%
\bibitem [{\citenamefont {Tacchino}\ \emph {et~al.}(2020)\citenamefont
  {Tacchino}, \citenamefont {Chiesa}, \citenamefont {Carretta},\ and\
  \citenamefont {Gerace}}]{Tacchino2020}%
  \BibitemOpen
  \bibfield  {author} {\bibinfo {author} {\bibfnamefont {F.}~\bibnamefont
  {Tacchino}}, \bibinfo {author} {\bibfnamefont {A.}~\bibnamefont {Chiesa}},
  \bibinfo {author} {\bibfnamefont {S.}~\bibnamefont {Carretta}},\ and\
  \bibinfo {author} {\bibfnamefont {D.}~\bibnamefont {Gerace}},\ }\bibfield
  {title} {\bibinfo {title} {Quantum computers as universal quantum simulators:
  State-of-the-art and perspectives},\ }\href
  {https://doi.org/https://doi.org/10.1002/qute.201900052} {\bibfield
  {journal} {\bibinfo  {journal} {Adv. Quant. Tech.}\ }\textbf {\bibinfo
  {volume} {3}},\ \bibinfo {pages} {1900052} (\bibinfo {year}
  {2020})}\BibitemShut {NoStop}%
\bibitem [{\citenamefont {{Van Dyke}}\ \emph {et~al.}(2022)\citenamefont {{Van
  Dyke}}, \citenamefont {Barnes}, \citenamefont {Economou},\ and\ \citenamefont
  {Nepomechie}}]{VanDyke2022}%
  \BibitemOpen
  \bibfield  {author} {\bibinfo {author} {\bibfnamefont {J.~S.}\ \bibnamefont
  {{Van Dyke}}}, \bibinfo {author} {\bibfnamefont {E.}~\bibnamefont {Barnes}},
  \bibinfo {author} {\bibfnamefont {S.~E.}\ \bibnamefont {Economou}},\ and\
  \bibinfo {author} {\bibfnamefont {R.~I.}\ \bibnamefont {Nepomechie}},\
  }\bibfield  {title} {\bibinfo {title} {Preparing exact eigenstates of the
  open {XXZ} chain on a quantum computer},\ }\href
  {https://doi.org/10.1088/1751-8121/ac4640} {\bibfield  {journal} {\bibinfo
  {journal} {J. Phys. A: Math. Theor.}\ }\textbf {\bibinfo {volume} {55}},\
  \bibinfo {pages} {055301} (\bibinfo {year} {2022})}\BibitemShut {NoStop}%
\bibitem [{\citenamefont {Khajetoorians}\ \emph {et~al.}(2011)\citenamefont
  {Khajetoorians}, \citenamefont {Wiebe}, \citenamefont {Chilian},\ and\
  \citenamefont {Wiesendanger}}]{Khajetoorians2011}%
  \BibitemOpen
  \bibfield  {author} {\bibinfo {author} {\bibfnamefont {A.~A.}\ \bibnamefont
  {Khajetoorians}}, \bibinfo {author} {\bibfnamefont {J.}~\bibnamefont
  {Wiebe}}, \bibinfo {author} {\bibfnamefont {B.}~\bibnamefont {Chilian}},\
  and\ \bibinfo {author} {\bibfnamefont {R.}~\bibnamefont {Wiesendanger}},\
  }\bibfield  {title} {\bibinfo {title} {Realizing all-spin\textendash based
  logic operations atom by atom},\ }\href
  {https://doi.org/10.1126/science.1201725} {\bibfield  {journal} {\bibinfo
  {journal} {Science}\ }\textbf {\bibinfo {volume} {332}},\ \bibinfo {pages}
  {1062} (\bibinfo {year} {2011})}\BibitemShut {NoStop}%
\bibitem [{\citenamefont {Khajetoorians}\ \emph {et~al.}(2013)\citenamefont
  {Khajetoorians}, \citenamefont {Baxevanis}, \citenamefont {Hübner},
  \citenamefont {Schlenk}, \citenamefont {Krause}, \citenamefont {Wehling},
  \citenamefont {Lounis}, \citenamefont {Lichtenstein}, \citenamefont
  {Pfannkuche}, \citenamefont {Wiebe},\ and\ \citenamefont
  {Wiesendanger}}]{Khajetoorians2013}%
  \BibitemOpen
  \bibfield  {author} {\bibinfo {author} {\bibfnamefont {A.~A.}\ \bibnamefont
  {Khajetoorians}}, \bibinfo {author} {\bibfnamefont {B.}~\bibnamefont
  {Baxevanis}}, \bibinfo {author} {\bibfnamefont {C.}~\bibnamefont {Hübner}},
  \bibinfo {author} {\bibfnamefont {T.}~\bibnamefont {Schlenk}}, \bibinfo
  {author} {\bibfnamefont {S.}~\bibnamefont {Krause}}, \bibinfo {author}
  {\bibfnamefont {T.~O.}\ \bibnamefont {Wehling}}, \bibinfo {author}
  {\bibfnamefont {S.}~\bibnamefont {Lounis}}, \bibinfo {author} {\bibfnamefont
  {A.}~\bibnamefont {Lichtenstein}}, \bibinfo {author} {\bibfnamefont
  {D.}~\bibnamefont {Pfannkuche}}, \bibinfo {author} {\bibfnamefont
  {J.}~\bibnamefont {Wiebe}},\ and\ \bibinfo {author} {\bibfnamefont
  {R.}~\bibnamefont {Wiesendanger}},\ }\bibfield  {title} {\bibinfo {title}
  {Current-driven spin dynamics of artificially constructed quantum magnets},\
  }\href {https://doi.org/10.1126/science.1228519} {\bibfield  {journal}
  {\bibinfo  {journal} {Science}\ }\textbf {\bibinfo {volume} {339}},\ \bibinfo
  {pages} {55} (\bibinfo {year} {2013})}\BibitemShut {NoStop}%
\bibitem [{\citenamefont {Chen}\ \emph
  {et~al.}(2023{\natexlab{b}})\citenamefont {Chen}, \citenamefont {Bae},\ and\
  \citenamefont {Heinrich}}]{Chen2022}%
  \BibitemOpen
  \bibfield  {author} {\bibinfo {author} {\bibfnamefont {Y.}~\bibnamefont
  {Chen}}, \bibinfo {author} {\bibfnamefont {Y.}~\bibnamefont {Bae}},\ and\
  \bibinfo {author} {\bibfnamefont {A.~J.}\ \bibnamefont {Heinrich}},\
  }\bibfield  {title} {\bibinfo {title} {Harnessing the quantum behavior of
  spins on surfaces},\ }\href
  {https://doi.org/https://doi.org/10.1002/adma.202107534} {\bibfield
  {journal} {\bibinfo  {journal} {Adv. Mater.}\ }\textbf {\bibinfo {volume}
  {35}},\ \bibinfo {pages} {2107534} (\bibinfo {year}
  {2023}{\natexlab{b}})}\BibitemShut {NoStop}%
\bibitem [{\citenamefont {Korepin}\ \emph {et~al.}(1993)\citenamefont
  {Korepin}, \citenamefont {Bogoliubov},\ and\ \citenamefont
  {Izergin}}]{Korepin1993}%
  \BibitemOpen
  \bibfield  {author} {\bibinfo {author} {\bibfnamefont {V.~E.}\ \bibnamefont
  {Korepin}}, \bibinfo {author} {\bibfnamefont {N.~M.}\ \bibnamefont
  {Bogoliubov}},\ and\ \bibinfo {author} {\bibfnamefont {A.~G.}\ \bibnamefont
  {Izergin}},\ }\href@noop {} {\emph {\bibinfo {title} {Quantum Inverse
  Scattering Method and Correlation Functions}}}\ (\bibinfo  {publisher}
  {Cambridge University Press, Cambridge},\ \bibinfo {year} {1993})\BibitemShut
  {NoStop}%
\bibitem [{\citenamefont {Jordan}\ and\ \citenamefont
  {Wigner}(1928)}]{Jordan1928}%
  \BibitemOpen
  \bibfield  {author} {\bibinfo {author} {\bibfnamefont {P.}~\bibnamefont
  {Jordan}}\ and\ \bibinfo {author} {\bibfnamefont {E.}~\bibnamefont
  {Wigner}},\ }\bibfield  {title} {\bibinfo {title} {{\"U}ber das {P}aulische
  {{\"A}}quivalenzverbot},\ }\href {https://doi.org/10.1007/BF01331938}
  {\bibfield  {journal} {\bibinfo  {journal} {Z. Phys.}\ }\textbf {\bibinfo
  {volume} {47}},\ \bibinfo {pages} {631} (\bibinfo {year} {1928})}\BibitemShut
  {NoStop}%
\bibitem [{\citenamefont {Lieb}\ \emph {et~al.}(1961)\citenamefont {Lieb},
  \citenamefont {Schultz},\ and\ \citenamefont {Mattis}}]{Lieb1961}%
  \BibitemOpen
  \bibfield  {author} {\bibinfo {author} {\bibfnamefont {E.}~\bibnamefont
  {Lieb}}, \bibinfo {author} {\bibfnamefont {T.}~\bibnamefont {Schultz}},\ and\
  \bibinfo {author} {\bibfnamefont {D.}~\bibnamefont {Mattis}},\ }\bibfield
  {title} {\bibinfo {title} {Two soluble models of an antiferromagnetic
  chain},\ }\href
  {https://doi.org/https://doi.org/10.1016/0003-4916(61)90115-4} {\bibfield
  {journal} {\bibinfo  {journal} {Ann. Phys.}\ }\textbf {\bibinfo {volume}
  {16}},\ \bibinfo {pages} {407} (\bibinfo {year} {1961})}\BibitemShut
  {NoStop}%
\bibitem [{\citenamefont {Kitaev}(2006)}]{Kitaev2006}%
  \BibitemOpen
  \bibfield  {author} {\bibinfo {author} {\bibfnamefont {A.}~\bibnamefont
  {Kitaev}},\ }\bibfield  {title} {\bibinfo {title} {Anyons in an exactly
  solved model and beyond},\ }\href
  {https://doi.org/https://doi.org/10.1016/j.aop.2005.10.005} {\bibfield
  {journal} {\bibinfo  {journal} {Ann. Phys.}\ }\textbf {\bibinfo {volume}
  {321}},\ \bibinfo {pages} {2} (\bibinfo {year} {2006})}\BibitemShut {NoStop}%
\bibitem [{\citenamefont {Wen}(2007)}]{Wen2007}%
  \BibitemOpen
  \bibfield  {author} {\bibinfo {author} {\bibfnamefont {X.-G.}\ \bibnamefont
  {Wen}},\ }\href@noop {} {\emph {\bibinfo {title} {{Quantum Field Theory of
  Many-Body Systems: From the Origin of Sound to an Origin of Light and
  Electrons}}}}\ (\bibinfo  {publisher} {Oxford University Press},\ \bibinfo
  {year} {Oxford, 2007})\BibitemShut {NoStop}%
\bibitem [{\citenamefont {Moudgalya}\ \emph {et~al.}(2022)\citenamefont
  {Moudgalya}, \citenamefont {Bernevig},\ and\ \citenamefont
  {Regnault}}]{Moudgalya2022}%
  \BibitemOpen
  \bibfield  {author} {\bibinfo {author} {\bibfnamefont {S.}~\bibnamefont
  {Moudgalya}}, \bibinfo {author} {\bibfnamefont {B.~A.}\ \bibnamefont
  {Bernevig}},\ and\ \bibinfo {author} {\bibfnamefont {N.}~\bibnamefont
  {Regnault}},\ }\bibfield  {title} {\bibinfo {title} {Quantum many-body scars
  and {Hilbert} space fragmentation: a review of exact results},\ }\href
  {https://doi.org/10.1088/1361-6633/ac73a0} {\bibfield  {journal} {\bibinfo
  {journal} {Rep. Prog. Phys.}\ }\textbf {\bibinfo {volume} {85}},\ \bibinfo
  {pages} {086501} (\bibinfo {year} {2022})}\BibitemShut {NoStop}%
\bibitem [{\citenamefont {Deutsch}(1991)}]{Deutsch1991}%
  \BibitemOpen
  \bibfield  {author} {\bibinfo {author} {\bibfnamefont {J.~M.}\ \bibnamefont
  {Deutsch}},\ }\bibfield  {title} {\bibinfo {title} {Quantum statistical
  mechanics in a closed system},\ }\href
  {https://doi.org/10.1103/PhysRevA.43.2046} {\bibfield  {journal} {\bibinfo
  {journal} {Phys. Rev. A}\ }\textbf {\bibinfo {volume} {43}},\ \bibinfo
  {pages} {2046} (\bibinfo {year} {1991})}\BibitemShut {NoStop}%
\bibitem [{\citenamefont {Srednicki}(1994)}]{Srednicki1994}%
  \BibitemOpen
  \bibfield  {author} {\bibinfo {author} {\bibfnamefont {M.}~\bibnamefont
  {Srednicki}},\ }\bibfield  {title} {\bibinfo {title} {Chaos and quantum
  thermalization},\ }\href {https://doi.org/10.1103/PhysRevE.50.888} {\bibfield
   {journal} {\bibinfo  {journal} {Phys. Rev. E}\ }\textbf {\bibinfo {volume}
  {50}},\ \bibinfo {pages} {888} (\bibinfo {year} {1994})}\BibitemShut
  {NoStop}%
\bibitem [{\citenamefont {Cao}\ \emph {et~al.}(2003)\citenamefont {Cao},
  \citenamefont {Lin}, \citenamefont {Shi},\ and\ \citenamefont
  {Wang}}]{Cao2003}%
  \BibitemOpen
  \bibfield  {author} {\bibinfo {author} {\bibfnamefont {J.}~\bibnamefont
  {Cao}}, \bibinfo {author} {\bibfnamefont {H.~Q.}\ \bibnamefont {Lin}},
  \bibinfo {author} {\bibfnamefont {K.~J.}\ \bibnamefont {Shi}},\ and\ \bibinfo
  {author} {\bibfnamefont {Y.}~\bibnamefont {Wang}},\ }\bibfield  {title}
  {\bibinfo {title} {Exact solution of {XXZ} spin chain with unparallel
  boundary fields},\ }\href {https://doi.org/10.1016/S0550-3213(03)00372-9}
  {\bibfield  {journal} {\bibinfo  {journal} {Nucl. Phys. B}\ }\textbf
  {\bibinfo {volume} {663}},\ \bibinfo {pages} {487} (\bibinfo {year}
  {2003})}\BibitemShut {NoStop}%
\bibitem [{\citenamefont {Batista}(2009)}]{Batista2009}%
  \BibitemOpen
  \bibfield  {author} {\bibinfo {author} {\bibfnamefont {C.~D.}\ \bibnamefont
  {Batista}},\ }\bibfield  {title} {\bibinfo {title} {Canted spiral: An exact
  ground state of {XXZ} zigzag spin ladders},\ }\href
  {https://doi.org/10.1103/PhysRevB.80.180406} {\bibfield  {journal} {\bibinfo
  {journal} {Phys. Rev. B}\ }\textbf {\bibinfo {volume} {80}},\ \bibinfo
  {pages} {180406} (\bibinfo {year} {2009})}\BibitemShut {NoStop}%
\bibitem [{\citenamefont {Batista}\ and\ \citenamefont
  {Somma}(2012)}]{Batista2012}%
  \BibitemOpen
  \bibfield  {author} {\bibinfo {author} {\bibfnamefont {C.~D.}\ \bibnamefont
  {Batista}}\ and\ \bibinfo {author} {\bibfnamefont {R.~D.}\ \bibnamefont
  {Somma}},\ }\bibfield  {title} {\bibinfo {title} {Condensation of anyons in
  frustrated quantum magnets},\ }\href
  {https://doi.org/10.1103/PhysRevLett.109.227203} {\bibfield  {journal}
  {\bibinfo  {journal} {Phys. Rev. Lett.}\ }\textbf {\bibinfo {volume} {109}},\
  \bibinfo {pages} {227203} (\bibinfo {year} {2012})}\BibitemShut {NoStop}%
\bibitem [{\citenamefont {Cerezo}\ \emph {et~al.}(2016)\citenamefont {Cerezo},
  \citenamefont {Rossignoli},\ and\ \citenamefont {Canosa}}]{Cerezo2016}%
  \BibitemOpen
  \bibfield  {author} {\bibinfo {author} {\bibfnamefont {M.}~\bibnamefont
  {Cerezo}}, \bibinfo {author} {\bibfnamefont {R.}~\bibnamefont {Rossignoli}},\
  and\ \bibinfo {author} {\bibfnamefont {N.}~\bibnamefont {Canosa}},\
  }\bibfield  {title} {\bibinfo {title} {Factorization in spin systems under
  general fields and separable ground-state engineering},\ }\href
  {https://doi.org/10.1103/PHYSREVA.94.042335/FIGURES/7/MEDIUM} {\bibfield
  {journal} {\bibinfo  {journal} {Phys. Rev. A}\ }\textbf {\bibinfo {volume}
  {94}},\ \bibinfo {pages} {042335} (\bibinfo {year} {2016})}\BibitemShut
  {NoStop}%
\bibitem [{\citenamefont {Popkov}\ \emph {et~al.}(2021)\citenamefont {Popkov},
  \citenamefont {Zhang},\ and\ \citenamefont {Kl\"umper}}]{Popkov2021}%
  \BibitemOpen
  \bibfield  {author} {\bibinfo {author} {\bibfnamefont {V.}~\bibnamefont
  {Popkov}}, \bibinfo {author} {\bibfnamefont {X.}~\bibnamefont {Zhang}},\ and\
  \bibinfo {author} {\bibfnamefont {A.}~\bibnamefont {Kl\"umper}},\ }\bibfield
  {title} {\bibinfo {title} {Phantom {B}ethe excitations and spin helix
  eigenstates in integrable periodic and open spin chains},\ }\href
  {https://doi.org/10.1103/PhysRevB.104.L081410} {\bibfield  {journal}
  {\bibinfo  {journal} {Phys. Rev. B}\ }\textbf {\bibinfo {volume} {104}},\
  \bibinfo {pages} {L081410} (\bibinfo {year} {2021})}\BibitemShut {NoStop}%
\bibitem [{\citenamefont {Zhang}\ \emph
  {et~al.}(2021{\natexlab{a}})\citenamefont {Zhang}, \citenamefont
  {Kl\"umper},\ and\ \citenamefont {Popkov}}]{Zhang2021-1}%
  \BibitemOpen
  \bibfield  {author} {\bibinfo {author} {\bibfnamefont {X.}~\bibnamefont
  {Zhang}}, \bibinfo {author} {\bibfnamefont {A.}~\bibnamefont {Kl\"umper}},\
  and\ \bibinfo {author} {\bibfnamefont {V.}~\bibnamefont {Popkov}},\
  }\bibfield  {title} {\bibinfo {title} {Phantom {B}ethe roots in the
  integrable open spin-$\frac{1}{2}$ {XXZ} chain},\ }\href
  {https://doi.org/10.1103/PhysRevB.103.115435} {\bibfield  {journal} {\bibinfo
   {journal} {Phys. Rev. B}\ }\textbf {\bibinfo {volume} {103}},\ \bibinfo
  {pages} {115435} (\bibinfo {year} {2021}{\natexlab{a}})}\BibitemShut
  {NoStop}%
\bibitem [{\citenamefont {Zhang}\ \emph
  {et~al.}(2021{\natexlab{b}})\citenamefont {Zhang}, \citenamefont
  {Kl\"umper},\ and\ \citenamefont {Popkov}}]{Zhang2021-2}%
  \BibitemOpen
  \bibfield  {author} {\bibinfo {author} {\bibfnamefont {X.}~\bibnamefont
  {Zhang}}, \bibinfo {author} {\bibfnamefont {A.}~\bibnamefont {Kl\"umper}},\
  and\ \bibinfo {author} {\bibfnamefont {V.}~\bibnamefont {Popkov}},\
  }\bibfield  {title} {\bibinfo {title} {Chiral coordinate {B}ethe ansatz for
  phantom eigenstates in the open {XXZ} spin-$\frac{1}{2}$ chain},\ }\href
  {https://doi.org/10.1103/PhysRevB.104.195409} {\bibfield  {journal} {\bibinfo
   {journal} {Phys. Rev. B}\ }\textbf {\bibinfo {volume} {104}},\ \bibinfo
  {pages} {195409} (\bibinfo {year} {2021}{\natexlab{b}})}\BibitemShut
  {NoStop}%
\bibitem [{\citenamefont {Zhang}\ \emph {et~al.}(2022)\citenamefont {Zhang},
  \citenamefont {Kl\"umper},\ and\ \citenamefont {Popkov}}]{Zhang2022}%
  \BibitemOpen
  \bibfield  {author} {\bibinfo {author} {\bibfnamefont {X.}~\bibnamefont
  {Zhang}}, \bibinfo {author} {\bibfnamefont {A.}~\bibnamefont {Kl\"umper}},\
  and\ \bibinfo {author} {\bibfnamefont {V.}~\bibnamefont {Popkov}},\
  }\bibfield  {title} {\bibinfo {title} {Invariant subspaces and elliptic
  spin-helix states in the integrable open spin-$\frac{1}{2}$ {XYZ} chain},\
  }\href {https://doi.org/10.1103/PhysRevB.106.075406} {\bibfield  {journal}
  {\bibinfo  {journal} {Phys. Rev. B}\ }\textbf {\bibinfo {volume} {106}},\
  \bibinfo {pages} {075406} (\bibinfo {year} {2022})}\BibitemShut {NoStop}%
\bibitem [{\citenamefont {Ma}\ \emph {et~al.}(2022)\citenamefont {Ma},
  \citenamefont {Zhang},\ and\ \citenamefont {Song}}]{Ma2022}%
  \BibitemOpen
  \bibfield  {author} {\bibinfo {author} {\bibfnamefont {E.~S.}\ \bibnamefont
  {Ma}}, \bibinfo {author} {\bibfnamefont {K.~L.}\ \bibnamefont {Zhang}},\ and\
  \bibinfo {author} {\bibfnamefont {Z.}~\bibnamefont {Song}},\ }\bibfield
  {title} {\bibinfo {title} {Steady helix states in a resonant {XXZ}
  {H}eisenberg model with {D}zyaloshinskii-{M}oriya interaction},\ }\href
  {https://doi.org/10.1103/PhysRevB.106.245122} {\bibfield  {journal} {\bibinfo
   {journal} {Phys. Rev. B}\ }\textbf {\bibinfo {volume} {106}},\ \bibinfo
  {pages} {245122} (\bibinfo {year} {2022})}\BibitemShut {NoStop}%
\bibitem [{\citenamefont {Popkov}\ \emph {et~al.}(2023)\citenamefont {Popkov},
  \citenamefont {\ifmmode \check{Z}\else
  \v{Z}\fi{}nidari\ifmmode~\check{c}\else \v{c}\fi{}},\ and\ \citenamefont
  {Zhang}}]{Popkov2023}%
  \BibitemOpen
  \bibfield  {author} {\bibinfo {author} {\bibfnamefont {V.}~\bibnamefont
  {Popkov}}, \bibinfo {author} {\bibfnamefont {M.}~\bibnamefont {\ifmmode
  \check{Z}\else \v{Z}\fi{}nidari\ifmmode~\check{c}\else \v{c}\fi{}}},\ and\
  \bibinfo {author} {\bibfnamefont {X.}~\bibnamefont {Zhang}},\ }\bibfield
  {title} {\bibinfo {title} {Universality in the relaxation of spin helices
  under {XXZ} spin-chain dynamics},\ }\href
  {https://doi.org/10.1103/PhysRevB.107.235408} {\bibfield  {journal} {\bibinfo
   {journal} {Phys. Rev. B}\ }\textbf {\bibinfo {volume} {107}},\ \bibinfo
  {pages} {235408} (\bibinfo {year} {2023})}\BibitemShut {NoStop}%
\bibitem [{\citenamefont {K\"uhn}\ \emph {et~al.}(2023)\citenamefont {K\"uhn},
  \citenamefont {Gerken}, \citenamefont {Funcke}, \citenamefont {Hartung},
  \citenamefont {Stornati}, \citenamefont {Jansen},\ and\ \citenamefont
  {Posske}}]{Kuehn2023}%
  \BibitemOpen
  \bibfield  {author} {\bibinfo {author} {\bibfnamefont {S.}~\bibnamefont
  {K\"uhn}}, \bibinfo {author} {\bibfnamefont {F.}~\bibnamefont {Gerken}},
  \bibinfo {author} {\bibfnamefont {L.}~\bibnamefont {Funcke}}, \bibinfo
  {author} {\bibfnamefont {T.}~\bibnamefont {Hartung}}, \bibinfo {author}
  {\bibfnamefont {P.}~\bibnamefont {Stornati}}, \bibinfo {author}
  {\bibfnamefont {K.}~\bibnamefont {Jansen}},\ and\ \bibinfo {author}
  {\bibfnamefont {T.}~\bibnamefont {Posske}},\ }\bibfield  {title} {\bibinfo
  {title} {Quantum spin helices more stable than the ground state: Onset of
  helical protection},\ }\href {https://doi.org/10.1103/PhysRevB.107.214422}
  {\bibfield  {journal} {\bibinfo  {journal} {Phys. Rev. B}\ }\textbf {\bibinfo
  {volume} {107}},\ \bibinfo {pages} {214422} (\bibinfo {year}
  {2023})}\BibitemShut {NoStop}%
\bibitem [{\citenamefont {Shi}\ and\ \citenamefont {Song}(2023)}]{Shi2023}%
  \BibitemOpen
  \bibfield  {author} {\bibinfo {author} {\bibfnamefont {Y.~B.}\ \bibnamefont
  {Shi}}\ and\ \bibinfo {author} {\bibfnamefont {Z.}~\bibnamefont {Song}},\
  }\bibfield  {title} {\bibinfo {title} {Robust unidirectional phantom helix
  states in the {XXZ} heisenberg model with {D}zyaloshinskii-{M}oriya
  interaction},\ }\href {https://doi.org/10.1103/PhysRevB.108.085108}
  {\bibfield  {journal} {\bibinfo  {journal} {Phys. Rev. B}\ }\textbf {\bibinfo
  {volume} {108}},\ \bibinfo {pages} {085108} (\bibinfo {year}
  {2023})}\BibitemShut {NoStop}%
\bibitem [{\citenamefont {Cerezo}\ \emph {et~al.}(2015)\citenamefont {Cerezo},
  \citenamefont {Rossignoli},\ and\ \citenamefont {Canosa}}]{Cerezo2015}%
  \BibitemOpen
  \bibfield  {author} {\bibinfo {author} {\bibfnamefont {M.}~\bibnamefont
  {Cerezo}}, \bibinfo {author} {\bibfnamefont {R.}~\bibnamefont {Rossignoli}},\
  and\ \bibinfo {author} {\bibfnamefont {N.}~\bibnamefont {Canosa}},\
  }\bibfield  {title} {\bibinfo {title} {Nontransverse factorizing fields and
  entanglement in finite spin systems},\ }\href
  {https://doi.org/10.1103/PHYSREVB.92.224422/FIGURES/7/MEDIUM} {\bibfield
  {journal} {\bibinfo  {journal} {Phys. Rev. B}\ }\textbf {\bibinfo {volume}
  {92}},\ \bibinfo {pages} {224422} (\bibinfo {year} {2015})}\BibitemShut
  {NoStop}%
\bibitem [{\citenamefont {Cerezo}\ \emph {et~al.}(2017)\citenamefont {Cerezo},
  \citenamefont {Rossignoli}, \citenamefont {Canosa},\ and\ \citenamefont
  {Ríos}}]{Cerezo2017}%
  \BibitemOpen
  \bibfield  {author} {\bibinfo {author} {\bibfnamefont {M.}~\bibnamefont
  {Cerezo}}, \bibinfo {author} {\bibfnamefont {R.}~\bibnamefont {Rossignoli}},
  \bibinfo {author} {\bibfnamefont {N.}~\bibnamefont {Canosa}},\ and\ \bibinfo
  {author} {\bibfnamefont {E.}~\bibnamefont {Ríos}},\ }\bibfield  {title}
  {\bibinfo {title} {Factorization and criticality in finite {XXZ} systems of
  arbitrary spin},\ }\href
  {https://doi.org/10.1103/PHYSREVLETT.119.220605/FIGURES/4/MEDIUM} {\bibfield
  {journal} {\bibinfo  {journal} {Phys. Rev. Lett.}\ }\textbf {\bibinfo
  {volume} {119}},\ \bibinfo {pages} {220605} (\bibinfo {year}
  {2017})}\BibitemShut {NoStop}%
\bibitem [{\citenamefont {Lee}\ \emph {et~al.}(2020)\citenamefont {Lee},
  \citenamefont {Melendrez}, \citenamefont {Pal},\ and\ \citenamefont
  {Changlani}}]{Lee2020}%
  \BibitemOpen
  \bibfield  {author} {\bibinfo {author} {\bibfnamefont {K.}~\bibnamefont
  {Lee}}, \bibinfo {author} {\bibfnamefont {R.}~\bibnamefont {Melendrez}},
  \bibinfo {author} {\bibfnamefont {A.}~\bibnamefont {Pal}},\ and\ \bibinfo
  {author} {\bibfnamefont {H.~J.}\ \bibnamefont {Changlani}},\ }\bibfield
  {title} {\bibinfo {title} {Exact three-colored quantum scars from geometric
  frustration},\ }\href {https://doi.org/10.1103/PhysRevB.101.241111}
  {\bibfield  {journal} {\bibinfo  {journal} {Phys. Rev. B}\ }\textbf {\bibinfo
  {volume} {101}},\ \bibinfo {pages} {241111} (\bibinfo {year}
  {2020})}\BibitemShut {NoStop}%
\bibitem [{\citenamefont {Changlani}\ \emph {et~al.}(2018)\citenamefont
  {Changlani}, \citenamefont {Kochkov}, \citenamefont {Kumar}, \citenamefont
  {Clark},\ and\ \citenamefont {Fradkin}}]{Changlani2018}%
  \BibitemOpen
  \bibfield  {author} {\bibinfo {author} {\bibfnamefont {H.~J.}\ \bibnamefont
  {Changlani}}, \bibinfo {author} {\bibfnamefont {D.}~\bibnamefont {Kochkov}},
  \bibinfo {author} {\bibfnamefont {K.}~\bibnamefont {Kumar}}, \bibinfo
  {author} {\bibfnamefont {B.~K.}\ \bibnamefont {Clark}},\ and\ \bibinfo
  {author} {\bibfnamefont {E.}~\bibnamefont {Fradkin}},\ }\bibfield  {title}
  {\bibinfo {title} {Macroscopically degenerate exactly solvable point in the
  spin-$1/2$ kagome quantum antiferromagnet},\ }\href
  {https://doi.org/10.1103/PhysRevLett.120.117202} {\bibfield  {journal}
  {\bibinfo  {journal} {Phys. Rev. Lett.}\ }\textbf {\bibinfo {volume} {120}},\
  \bibinfo {pages} {117202} (\bibinfo {year} {2018})}\BibitemShut {NoStop}%
\bibitem [{\citenamefont {Pal}\ \emph {et~al.}(2021)\citenamefont {Pal},
  \citenamefont {Sharma}, \citenamefont {Changlani},\ and\ \citenamefont
  {Pujari}}]{Pal2021}%
  \BibitemOpen
  \bibfield  {author} {\bibinfo {author} {\bibfnamefont {S.}~\bibnamefont
  {Pal}}, \bibinfo {author} {\bibfnamefont {P.}~\bibnamefont {Sharma}},
  \bibinfo {author} {\bibfnamefont {H.~J.}\ \bibnamefont {Changlani}},\ and\
  \bibinfo {author} {\bibfnamefont {S.}~\bibnamefont {Pujari}},\ }\bibfield
  {title} {\bibinfo {title} {Colorful points in the {XY} regime of {XXZ}
  quantum magnets},\ }\href {https://doi.org/10.1103/PhysRevB.103.144414}
  {\bibfield  {journal} {\bibinfo  {journal} {Phys. Rev. B}\ }\textbf {\bibinfo
  {volume} {103}},\ \bibinfo {pages} {144414} (\bibinfo {year}
  {2021})}\BibitemShut {NoStop}%
\bibitem [{\citenamefont {Chertkov}\ and\ \citenamefont
  {Clark}(2021)}]{Chertkov2021}%
  \BibitemOpen
  \bibfield  {author} {\bibinfo {author} {\bibfnamefont {E.}~\bibnamefont
  {Chertkov}}\ and\ \bibinfo {author} {\bibfnamefont {B.~K.}\ \bibnamefont
  {Clark}},\ }\bibfield  {title} {\bibinfo {title} {Motif magnetism and quantum
  many-body scars},\ }\href {https://doi.org/10.1103/PhysRevB.104.104410}
  {\bibfield  {journal} {\bibinfo  {journal} {Phys. Rev. B}\ }\textbf {\bibinfo
  {volume} {104}},\ \bibinfo {pages} {104410} (\bibinfo {year}
  {2021})}\BibitemShut {NoStop}%
\bibitem [{\citenamefont {Kim}\ and\ \citenamefont
  {Tserkovnyak}(2016)}]{Kim2016}%
  \BibitemOpen
  \bibfield  {author} {\bibinfo {author} {\bibfnamefont {S.~K.}\ \bibnamefont
  {Kim}}\ and\ \bibinfo {author} {\bibfnamefont {Y.}~\bibnamefont
  {Tserkovnyak}},\ }\bibfield  {title} {\bibinfo {title} {Topological effects
  on quantum phase slips in superfluid spin transport},\ }\href
  {https://doi.org/10.1103/PhysRevLett.116.127201} {\bibfield  {journal}
  {\bibinfo  {journal} {Phys. Rev. Lett.}\ }\textbf {\bibinfo {volume} {116}},\
  \bibinfo {pages} {127201} (\bibinfo {year} {2016})}\BibitemShut {NoStop}%
\bibitem [{\citenamefont {Tserkovnyak}\ \emph {et~al.}(2020)\citenamefont
  {Tserkovnyak}, \citenamefont {Zou}, \citenamefont {Kim},\ and\ \citenamefont
  {Takei}}]{Tserkovnyak2020}%
  \BibitemOpen
  \bibfield  {author} {\bibinfo {author} {\bibfnamefont {Y.}~\bibnamefont
  {Tserkovnyak}}, \bibinfo {author} {\bibfnamefont {J.}~\bibnamefont {Zou}},
  \bibinfo {author} {\bibfnamefont {S.~K.}\ \bibnamefont {Kim}},\ and\ \bibinfo
  {author} {\bibfnamefont {S.}~\bibnamefont {Takei}},\ }\bibfield  {title}
  {\bibinfo {title} {Quantum hydrodynamics of spin winding},\ }\href
  {https://doi.org/10.1103/PhysRevB.102.224433} {\bibfield  {journal} {\bibinfo
   {journal} {Phys. Rev. B}\ }\textbf {\bibinfo {volume} {102}},\ \bibinfo
  {pages} {224433} (\bibinfo {year} {2020})}\BibitemShut {NoStop}%
\bibitem [{\citenamefont {Takashima}\ \emph {et~al.}(2016)\citenamefont
  {Takashima}, \citenamefont {Ishizuka},\ and\ \citenamefont
  {Balents}}]{Takashima2016}%
  \BibitemOpen
  \bibfield  {author} {\bibinfo {author} {\bibfnamefont {R.}~\bibnamefont
  {Takashima}}, \bibinfo {author} {\bibfnamefont {H.}~\bibnamefont
  {Ishizuka}},\ and\ \bibinfo {author} {\bibfnamefont {L.}~\bibnamefont
  {Balents}},\ }\bibfield  {title} {\bibinfo {title} {Quantum skyrmions in
  two-dimensional chiral magnets},\ }\href
  {https://doi.org/10.1103/PhysRevB.94.134415} {\bibfield  {journal} {\bibinfo
  {journal} {Phys. Rev. B}\ }\textbf {\bibinfo {volume} {94}},\ \bibinfo
  {pages} {134415} (\bibinfo {year} {2016})}\BibitemShut {NoStop}%
\bibitem [{\citenamefont {Sotnikov}\ \emph {et~al.}(2023)\citenamefont
  {Sotnikov}, \citenamefont {Stepanov}, \citenamefont {Katsnelson},
  \citenamefont {Mila},\ and\ \citenamefont {Mazurenko}}]{Sotnikov2022}%
  \BibitemOpen
  \bibfield  {author} {\bibinfo {author} {\bibfnamefont {O.~M.}\ \bibnamefont
  {Sotnikov}}, \bibinfo {author} {\bibfnamefont {E.~A.}\ \bibnamefont
  {Stepanov}}, \bibinfo {author} {\bibfnamefont {M.~I.}\ \bibnamefont
  {Katsnelson}}, \bibinfo {author} {\bibfnamefont {F.}~\bibnamefont {Mila}},\
  and\ \bibinfo {author} {\bibfnamefont {V.~V.}\ \bibnamefont {Mazurenko}},\
  }\bibfield  {title} {\bibinfo {title} {Emergence of classical magnetic order
  from {A}nderson towers: Quantum {D}arwinism in action},\ }\href
  {https://doi.org/10.1103/PhysRevX.13.041027} {\bibfield  {journal} {\bibinfo
  {journal} {Phys. Rev. X}\ }\textbf {\bibinfo {volume} {13}},\ \bibinfo
  {pages} {041027} (\bibinfo {year} {2023})}\BibitemShut {NoStop}%
\bibitem [{\citenamefont {Haller}\ \emph {et~al.}(2022)\citenamefont {Haller},
  \citenamefont {Groenendijk}, \citenamefont {Habibi}, \citenamefont
  {Michels},\ and\ \citenamefont {Schmidt}}]{Haller2022}%
  \BibitemOpen
  \bibfield  {author} {\bibinfo {author} {\bibfnamefont {A.}~\bibnamefont
  {Haller}}, \bibinfo {author} {\bibfnamefont {S.}~\bibnamefont {Groenendijk}},
  \bibinfo {author} {\bibfnamefont {A.}~\bibnamefont {Habibi}}, \bibinfo
  {author} {\bibfnamefont {A.}~\bibnamefont {Michels}},\ and\ \bibinfo {author}
  {\bibfnamefont {T.~L.}\ \bibnamefont {Schmidt}},\ }\bibfield  {title}
  {\bibinfo {title} {Quantum skyrmion lattices in {H}eisenberg ferromagnets},\
  }\href {https://doi.org/10.1103/PHYSREVRESEARCH.4.043113/FIGURES/13/MEDIUM}
  {\bibfield  {journal} {\bibinfo  {journal} {Phys. Rev. Res.}\ }\textbf
  {\bibinfo {volume} {4}},\ \bibinfo {pages} {043113} (\bibinfo {year}
  {2022})}\BibitemShut {NoStop}%
\bibitem [{\citenamefont {Joshi}\ \emph {et~al.}(2023)\citenamefont {Joshi},
  \citenamefont {Peters},\ and\ \citenamefont {Posske}}]{Joshi2023}%
  \BibitemOpen
  \bibfield  {author} {\bibinfo {author} {\bibfnamefont {A.}~\bibnamefont
  {Joshi}}, \bibinfo {author} {\bibfnamefont {R.}~\bibnamefont {Peters}},\ and\
  \bibinfo {author} {\bibfnamefont {T.}~\bibnamefont {Posske}},\ }\bibfield
  {title} {\bibinfo {title} {Ground state properties of quantum skyrmions
  described by neural network quantum states},\ }\href
  {https://doi.org/10.1103/PhysRevB.108.094410} {\bibfield  {journal} {\bibinfo
   {journal} {Phys. Rev. B}\ }\textbf {\bibinfo {volume} {108}},\ \bibinfo
  {pages} {094410} (\bibinfo {year} {2023})}\BibitemShut {NoStop}%
\bibitem [{\citenamefont {Palle}\ and\ \citenamefont
  {Benton}(2021)}]{Palle2021}%
  \BibitemOpen
  \bibfield  {author} {\bibinfo {author} {\bibfnamefont {G.}~\bibnamefont
  {Palle}}\ and\ \bibinfo {author} {\bibfnamefont {O.}~\bibnamefont {Benton}},\
  }\bibfield  {title} {\bibinfo {title} {Exactly solvable spin-$\frac{1}{2}$
  {XYZ} models with highly degenerate partially ordered ground states},\ }\href
  {https://doi.org/10.1103/PhysRevB.103.214428} {\bibfield  {journal} {\bibinfo
   {journal} {Phys. Rev. B}\ }\textbf {\bibinfo {volume} {103}},\ \bibinfo
  {pages} {214428} (\bibinfo {year} {2021})}\BibitemShut {NoStop}%
\bibitem [{\citenamefont {Dmitriev}\ and\ \citenamefont
  {Krivnov}(2021)}]{Dmitriev2021}%
  \BibitemOpen
  \bibfield  {author} {\bibinfo {author} {\bibfnamefont {D.~V.}\ \bibnamefont
  {Dmitriev}}\ and\ \bibinfo {author} {\bibfnamefont {V.~Y.}\ \bibnamefont
  {Krivnov}},\ }\bibfield  {title} {\bibinfo {title} {Two-dimensional spin
  models with macroscopic degeneracy},\ }\href
  {https://doi.org/10.1088/1361-648X/ac18f4} {\bibfield  {journal} {\bibinfo
  {journal} {J. Phys.: Cond. Mat.}\ }\textbf {\bibinfo {volume} {33}},\
  \bibinfo {pages} {435802} (\bibinfo {year} {2021})}\BibitemShut {NoStop}%
\bibitem [{\citenamefont {Nussinov}\ and\ \citenamefont
  {Ortiz}(2023)}]{Nussinov2023}%
  \BibitemOpen
  \bibfield  {author} {\bibinfo {author} {\bibfnamefont {Z.}~\bibnamefont
  {Nussinov}}\ and\ \bibinfo {author} {\bibfnamefont {G.}~\bibnamefont
  {Ortiz}},\ }\bibfield  {title} {\bibinfo {title} {Theorem on extensive
  spectral degeneracy for systems with rigid higher symmetries in general
  dimensions},\ }\href {https://doi.org/10.1103/PhysRevB.107.045109} {\bibfield
   {journal} {\bibinfo  {journal} {Phys. Rev. B}\ }\textbf {\bibinfo {volume}
  {107}},\ \bibinfo {pages} {045109} (\bibinfo {year} {2023})}\BibitemShut
  {NoStop}%
\bibitem [{\citenamefont {Fendley}(2019)}]{Fendley2019}%
  \BibitemOpen
  \bibfield  {author} {\bibinfo {author} {\bibfnamefont {P.}~\bibnamefont
  {Fendley}},\ }\bibfield  {title} {\bibinfo {title} {Free fermions in
  disguise},\ }\href {https://doi.org/10.1088/1751-8121/ab305d} {\bibfield
  {journal} {\bibinfo  {journal} {J. Phys. A: Math. Theor.}\ }\textbf {\bibinfo
  {volume} {52}},\ \bibinfo {pages} {335002} (\bibinfo {year}
  {2019})}\BibitemShut {NoStop}%
\bibitem [{\citenamefont {Zadnik}\ and\ \citenamefont
  {Fagotti}(2021)}]{Zadnik2021-1}%
  \BibitemOpen
  \bibfield  {author} {\bibinfo {author} {\bibfnamefont {L.}~\bibnamefont
  {Zadnik}}\ and\ \bibinfo {author} {\bibfnamefont {M.}~\bibnamefont
  {Fagotti}},\ }\bibfield  {title} {\bibinfo {title} {{The folded spin-1/2 XXZ
  model: I. Diagonalisation, jamming, and ground state properties}},\ }\href
  {https://doi.org/10.21468/SciPostPhysCore.4.2.010} {\bibfield  {journal}
  {\bibinfo  {journal} {SciPost Phys. Core}\ }\textbf {\bibinfo {volume} {4}},\
  \bibinfo {pages} {010} (\bibinfo {year} {2021})}\BibitemShut {NoStop}%
\bibitem [{\citenamefont {Zadnik}\ \emph {et~al.}(2021)\citenamefont {Zadnik},
  \citenamefont {Bidzhiev},\ and\ \citenamefont {Fagotti}}]{Zadnik2021-2}%
  \BibitemOpen
  \bibfield  {author} {\bibinfo {author} {\bibfnamefont {L.}~\bibnamefont
  {Zadnik}}, \bibinfo {author} {\bibfnamefont {K.}~\bibnamefont {Bidzhiev}},\
  and\ \bibinfo {author} {\bibfnamefont {M.}~\bibnamefont {Fagotti}},\
  }\bibfield  {title} {\bibinfo {title} {{The folded spin-1/2 XXZ model: II.
  Thermodynamics and hydrodynamics with a minimal set of charges}},\ }\href
  {https://doi.org/10.21468/SciPostPhys.10.5.099} {\bibfield  {journal}
  {\bibinfo  {journal} {SciPost Phys.}\ }\textbf {\bibinfo {volume} {10}},\
  \bibinfo {pages} {099} (\bibinfo {year} {2021})}\BibitemShut {NoStop}%
\bibitem [{\citenamefont {Pozsgay}\ \emph {et~al.}(2021)\citenamefont
  {Pozsgay}, \citenamefont {Gombor}, \citenamefont {Hutsalyuk}, \citenamefont
  {Jiang}, \citenamefont {Pristy\'ak},\ and\ \citenamefont
  {Vernier}}]{Pozsgay2021}%
  \BibitemOpen
  \bibfield  {author} {\bibinfo {author} {\bibfnamefont {B.}~\bibnamefont
  {Pozsgay}}, \bibinfo {author} {\bibfnamefont {T.}~\bibnamefont {Gombor}},
  \bibinfo {author} {\bibfnamefont {A.}~\bibnamefont {Hutsalyuk}}, \bibinfo
  {author} {\bibfnamefont {Y.}~\bibnamefont {Jiang}}, \bibinfo {author}
  {\bibfnamefont {L.}~\bibnamefont {Pristy\'ak}},\ and\ \bibinfo {author}
  {\bibfnamefont {E.}~\bibnamefont {Vernier}},\ }\bibfield  {title} {\bibinfo
  {title} {Integrable spin chain with {H}ilbert space fragmentation and
  solvable real-time dynamics},\ }\href
  {https://doi.org/10.1103/PhysRevE.104.044106} {\bibfield  {journal} {\bibinfo
   {journal} {Phys. Rev. E}\ }\textbf {\bibinfo {volume} {104}},\ \bibinfo
  {pages} {044106} (\bibinfo {year} {2021})}\BibitemShut {NoStop}%
\bibitem [{\citenamefont {Murgan}\ and\ \citenamefont
  {Nepomechie}(2005)}]{Murgan2005}%
  \BibitemOpen
  \bibfield  {author} {\bibinfo {author} {\bibfnamefont {R.}~\bibnamefont
  {Murgan}}\ and\ \bibinfo {author} {\bibfnamefont {R.~I.}\ \bibnamefont
  {Nepomechie}},\ }\bibfield  {title} {\bibinfo {title} {{Generalized $T$-$Q$
  relations and the open {XXZ} chain}},\ }\href
  {https://doi.org/10.1088/1742-5468/2005/08/P08002} {\bibfield  {journal}
  {\bibinfo  {journal} {J. Stat. Mech.}\ }\textbf {\bibinfo {volume} {0508}},\
  \bibinfo {pages} {P08002} (\bibinfo {year} {2005})}\BibitemShut {NoStop}%
\bibitem [{\citenamefont {Eckardt}\ \emph {et~al.}(2010)\citenamefont
  {Eckardt}, \citenamefont {Hauke}, \citenamefont {Soltan-Panahi},
  \citenamefont {Becker}, \citenamefont {Sengstock},\ and\ \citenamefont
  {Lewenstein}}]{Eckardt2010}%
  \BibitemOpen
  \bibfield  {author} {\bibinfo {author} {\bibfnamefont {A.}~\bibnamefont
  {Eckardt}}, \bibinfo {author} {\bibfnamefont {P.}~\bibnamefont {Hauke}},
  \bibinfo {author} {\bibfnamefont {P.}~\bibnamefont {Soltan-Panahi}}, \bibinfo
  {author} {\bibfnamefont {C.}~\bibnamefont {Becker}}, \bibinfo {author}
  {\bibfnamefont {K.}~\bibnamefont {Sengstock}},\ and\ \bibinfo {author}
  {\bibfnamefont {M.}~\bibnamefont {Lewenstein}},\ }\bibfield  {title}
  {\bibinfo {title} {Frustrated quantum antiferromagnetism with ultracold
  bosons in a triangular lattice},\ }\href
  {https://doi.org/10.1209/0295-5075/89/10010} {\bibfield  {journal} {\bibinfo
  {journal} {Europhys. Lett.}\ }\textbf {\bibinfo {volume} {89}},\ \bibinfo
  {pages} {10010} (\bibinfo {year} {2010})}\BibitemShut {NoStop}%
\bibitem [{\citenamefont {Struck}\ \emph {et~al.}(2011)\citenamefont {Struck},
  \citenamefont {Ölschläger}, \citenamefont {Targat}, \citenamefont
  {Soltan-Panahi}, \citenamefont {Eckardt}, \citenamefont {Lewenstein},
  \citenamefont {Windpassinger},\ and\ \citenamefont {Sengstock}}]{Struck2011}%
  \BibitemOpen
  \bibfield  {author} {\bibinfo {author} {\bibfnamefont {J.}~\bibnamefont
  {Struck}}, \bibinfo {author} {\bibfnamefont {C.}~\bibnamefont
  {Ölschläger}}, \bibinfo {author} {\bibfnamefont {R.~L.}\ \bibnamefont
  {Targat}}, \bibinfo {author} {\bibfnamefont {P.}~\bibnamefont
  {Soltan-Panahi}}, \bibinfo {author} {\bibfnamefont {A.}~\bibnamefont
  {Eckardt}}, \bibinfo {author} {\bibfnamefont {M.}~\bibnamefont {Lewenstein}},
  \bibinfo {author} {\bibfnamefont {P.}~\bibnamefont {Windpassinger}},\ and\
  \bibinfo {author} {\bibfnamefont {K.}~\bibnamefont {Sengstock}},\ }\bibfield
  {title} {\bibinfo {title} {Quantum simulation of frustrated classical
  magnetism in triangular optical lattices},\ }\href
  {https://doi.org/10.1126/science.1207239} {\bibfield  {journal} {\bibinfo
  {journal} {Science}\ }\textbf {\bibinfo {volume} {333}},\ \bibinfo {pages}
  {996} (\bibinfo {year} {2011})}\BibitemShut {NoStop}%
\bibitem [{\citenamefont {Struck}\ \emph {et~al.}(2013)\citenamefont {Struck},
  \citenamefont {Weinberg}, \citenamefont {{\"O}lschl{\"a}ger}, \citenamefont
  {Windpassinger}, \citenamefont {Simonet}, \citenamefont {Sengstock},
  \citenamefont {H{\"o}ppner}, \citenamefont {Hauke}, \citenamefont {Eckardt},
  \citenamefont {Lewenstein},\ and\ \citenamefont {Mathey}}]{Struck2013}%
  \BibitemOpen
  \bibfield  {author} {\bibinfo {author} {\bibfnamefont {J.}~\bibnamefont
  {Struck}}, \bibinfo {author} {\bibfnamefont {M.}~\bibnamefont {Weinberg}},
  \bibinfo {author} {\bibfnamefont {C.}~\bibnamefont {{\"O}lschl{\"a}ger}},
  \bibinfo {author} {\bibfnamefont {P.}~\bibnamefont {Windpassinger}}, \bibinfo
  {author} {\bibfnamefont {J.}~\bibnamefont {Simonet}}, \bibinfo {author}
  {\bibfnamefont {K.}~\bibnamefont {Sengstock}}, \bibinfo {author}
  {\bibfnamefont {R.}~\bibnamefont {H{\"o}ppner}}, \bibinfo {author}
  {\bibfnamefont {P.}~\bibnamefont {Hauke}}, \bibinfo {author} {\bibfnamefont
  {A.}~\bibnamefont {Eckardt}}, \bibinfo {author} {\bibfnamefont
  {M.}~\bibnamefont {Lewenstein}},\ and\ \bibinfo {author} {\bibfnamefont
  {L.}~\bibnamefont {Mathey}},\ }\bibfield  {title} {\bibinfo {title}
  {Engineering {I}sing-{XY} spin-models in a triangular lattice using tunable
  artificial gauge fields},\ }\href {https://doi.org/10.1038/nphys2750}
  {\bibfield  {journal} {\bibinfo  {journal} {Nat. Phys.}\ }\textbf {\bibinfo
  {volume} {9}},\ \bibinfo {pages} {738} (\bibinfo {year} {2013})}\BibitemShut
  {NoStop}%
\bibitem [{\citenamefont {Vorberg}\ \emph {et~al.}(2013)\citenamefont
  {Vorberg}, \citenamefont {Wustmann}, \citenamefont {Ketzmerick},\ and\
  \citenamefont {Eckardt}}]{Vorberg2013}%
  \BibitemOpen
  \bibfield  {author} {\bibinfo {author} {\bibfnamefont {D.}~\bibnamefont
  {Vorberg}}, \bibinfo {author} {\bibfnamefont {W.}~\bibnamefont {Wustmann}},
  \bibinfo {author} {\bibfnamefont {R.}~\bibnamefont {Ketzmerick}},\ and\
  \bibinfo {author} {\bibfnamefont {A.}~\bibnamefont {Eckardt}},\ }\bibfield
  {title} {\bibinfo {title} {Generalized {B}ose-{E}instein condensation into
  multiple states in driven-dissipative systems},\ }\href
  {https://doi.org/10.1103/PhysRevLett.111.240405} {\bibfield  {journal}
  {\bibinfo  {journal} {Phys. Rev. Lett.}\ }\textbf {\bibinfo {volume} {111}},\
  \bibinfo {pages} {240405} (\bibinfo {year} {2013})}\BibitemShut {NoStop}%
\bibitem [{\citenamefont {Wu}\ and\ \citenamefont {Eckardt}(2022)}]{Wu2022}%
  \BibitemOpen
  \bibfield  {author} {\bibinfo {author} {\bibfnamefont {L.-N.}\ \bibnamefont
  {Wu}}\ and\ \bibinfo {author} {\bibfnamefont {A.}~\bibnamefont {Eckardt}},\
  }\bibfield  {title} {\bibinfo {title} {Cooling and state preparation in an
  optical lattice via {M}arkovian feedback control},\ }\href
  {https://doi.org/10.1103/PhysRevResearch.4.L022045} {\bibfield  {journal}
  {\bibinfo  {journal} {Phys. Rev. Res.}\ }\textbf {\bibinfo {volume} {4}},\
  \bibinfo {pages} {L022045} (\bibinfo {year} {2022})}\BibitemShut {NoStop}%
\bibitem [{\citenamefont {Petiziol}\ and\ \citenamefont
  {Eckardt}(2024)}]{Petiziol2024}%
  \BibitemOpen
  \bibfield  {author} {\bibinfo {author} {\bibfnamefont {F.}~\bibnamefont
  {Petiziol}}\ and\ \bibinfo {author} {\bibfnamefont {A.}~\bibnamefont
  {Eckardt}},\ }\bibfield  {title} {\bibinfo {title} {Controlling
  nonequilibrium {B}ose-{E}instein condensation with engineered environments},\
  }\href {https://doi.org/10.1103/PhysRevA.110.L021701} {\bibfield  {journal}
  {\bibinfo  {journal} {Phys. Rev. A}\ }\textbf {\bibinfo {volume} {110}},\
  \bibinfo {pages} {L021701} (\bibinfo {year} {2024})}\BibitemShut {NoStop}%
\bibitem [{\citenamefont {Schnell}\ \emph {et~al.}(2023)\citenamefont
  {Schnell}, \citenamefont {Wu}, \citenamefont {Widera},\ and\ \citenamefont
  {Eckardt}}]{Schnell2023}%
  \BibitemOpen
  \bibfield  {author} {\bibinfo {author} {\bibfnamefont {A.}~\bibnamefont
  {Schnell}}, \bibinfo {author} {\bibfnamefont {L.-N.}\ \bibnamefont {Wu}},
  \bibinfo {author} {\bibfnamefont {A.}~\bibnamefont {Widera}},\ and\ \bibinfo
  {author} {\bibfnamefont {A.}~\bibnamefont {Eckardt}},\ }\bibfield  {title}
  {\bibinfo {title} {Floquet-heating-induced {B}ose condensation in a scarlike
  mode of an open driven optical-lattice system},\ }\href
  {https://doi.org/10.1103/PhysRevA.107.L021301} {\bibfield  {journal}
  {\bibinfo  {journal} {Phys. Rev. A}\ }\textbf {\bibinfo {volume} {107}},\
  \bibinfo {pages} {L021301} (\bibinfo {year} {2023})}\BibitemShut {NoStop}%
\bibitem [{\citenamefont {Zhang}\ \emph {et~al.}(2023)\citenamefont {Zhang},
  \citenamefont {Shi},\ and\ \citenamefont {Song}}]{Zhang2023}%
  \BibitemOpen
  \bibfield  {author} {\bibinfo {author} {\bibfnamefont {C.~H.}\ \bibnamefont
  {Zhang}}, \bibinfo {author} {\bibfnamefont {Y.~B.}\ \bibnamefont {Shi}},\
  and\ \bibinfo {author} {\bibfnamefont {Z.}~\bibnamefont {Song}},\ }\href@noop
  {} {\bibinfo {title} {Generalized phantom helix states in quantum spin
  graphs}} (\bibinfo {year} {2023}),\ \Eprint
  {https://arxiv.org/abs/2310.11786} {arXiv:2310.11786 [quant-ph]} \BibitemShut
  {NoStop}%
\bibitem [{Note1()}]{Note1}%
  \BibitemOpen
  \bibinfo {note} {See Supplemental Material at [URL will be inserted by the
  publisher] for details.}\BibitemShut {Stop}%
\bibitem [{\citenamefont {Perk}\ and\ \citenamefont {Capel}(1976)}]{Perk1976}%
  \BibitemOpen
  \bibfield  {author} {\bibinfo {author} {\bibfnamefont {J.}~\bibnamefont
  {Perk}}\ and\ \bibinfo {author} {\bibfnamefont {H.}~\bibnamefont {Capel}},\
  }\bibfield  {title} {\bibinfo {title} {Antisymmetric exchange, canting and
  spiral structure},\ }\href
  {https://doi.org/https://doi.org/10.1016/0375-9601(76)90515-6} {\bibfield
  {journal} {\bibinfo  {journal} {Phys. Lett. A}\ }\textbf {\bibinfo {volume}
  {58}},\ \bibinfo {pages} {115} (\bibinfo {year} {1976})}\BibitemShut
  {NoStop}%
\bibitem [{\citenamefont {Chandrasekharan}\ and\ \citenamefont
  {Wiese}(1997)}]{Chandrasekharan1997}%
  \BibitemOpen
  \bibfield  {author} {\bibinfo {author} {\bibfnamefont {S.}~\bibnamefont
  {Chandrasekharan}}\ and\ \bibinfo {author} {\bibfnamefont {U.-J.}\
  \bibnamefont {Wiese}},\ }\bibfield  {title} {\bibinfo {title} {Quantum link
  models: A discrete approach to gauge theories},\ }\href
  {https://doi.org/https://doi.org/10.1016/S0550-3213(97)80041-7} {\bibfield
  {journal} {\bibinfo  {journal} {Nuclear Physics B}\ }\textbf {\bibinfo
  {volume} {492}},\ \bibinfo {pages} {455} (\bibinfo {year}
  {1997})}\BibitemShut {NoStop}%
\bibitem [{\citenamefont {Kim}\ \emph {et~al.}(2013)\citenamefont {Kim},
  \citenamefont {Lee}, \citenamefont {Lee},\ and\ \citenamefont
  {Stiles}}]{Kim2013}%
  \BibitemOpen
  \bibfield  {author} {\bibinfo {author} {\bibfnamefont {K.-W.}\ \bibnamefont
  {Kim}}, \bibinfo {author} {\bibfnamefont {H.-W.}\ \bibnamefont {Lee}},
  \bibinfo {author} {\bibfnamefont {K.-J.}\ \bibnamefont {Lee}},\ and\ \bibinfo
  {author} {\bibfnamefont {M.~D.}\ \bibnamefont {Stiles}},\ }\bibfield  {title}
  {\bibinfo {title} {Chirality from interfacial spin-orbit coupling effects in
  magnetic bilayers},\ }\href {https://doi.org/10.1103/PhysRevLett.111.216601}
  {\bibfield  {journal} {\bibinfo  {journal} {Phys. Rev. Lett.}\ }\textbf
  {\bibinfo {volume} {111}},\ \bibinfo {pages} {216601} (\bibinfo {year}
  {2013})}\BibitemShut {NoStop}%
\bibitem [{\citenamefont {Shanavas}\ and\ \citenamefont
  {Satpathy}(2014)}]{Shanavas2014}%
  \BibitemOpen
  \bibfield  {author} {\bibinfo {author} {\bibfnamefont {K.~V.}\ \bibnamefont
  {Shanavas}}\ and\ \bibinfo {author} {\bibfnamefont {S.}~\bibnamefont
  {Satpathy}},\ }\bibfield  {title} {\bibinfo {title} {Electric field tuning of
  the {R}ashba effect in the polar perovskite structures},\ }\href
  {https://doi.org/10.1103/PhysRevLett.112.086802} {\bibfield  {journal}
  {\bibinfo  {journal} {Phys. Rev. Lett.}\ }\textbf {\bibinfo {volume} {112}},\
  \bibinfo {pages} {086802} (\bibinfo {year} {2014})}\BibitemShut {NoStop}%
\bibitem [{\citenamefont {Zhang}\ \emph {et~al.}(2018)\citenamefont {Zhang},
  \citenamefont {Zhong}, \citenamefont {Zang}, \citenamefont {Zhang},
  \citenamefont {Yu}, \citenamefont {Han}, \citenamefont {Liu}, \citenamefont
  {Yan}, \citenamefont {Kang},\ and\ \citenamefont {Mei}}]{Zhang2018}%
  \BibitemOpen
  \bibfield  {author} {\bibinfo {author} {\bibfnamefont {W.}~\bibnamefont
  {Zhang}}, \bibinfo {author} {\bibfnamefont {H.}~\bibnamefont {Zhong}},
  \bibinfo {author} {\bibfnamefont {R.}~\bibnamefont {Zang}}, \bibinfo {author}
  {\bibfnamefont {Y.}~\bibnamefont {Zhang}}, \bibinfo {author} {\bibfnamefont
  {S.}~\bibnamefont {Yu}}, \bibinfo {author} {\bibfnamefont {G.}~\bibnamefont
  {Han}}, \bibinfo {author} {\bibfnamefont {G.}~\bibnamefont {Liu}}, \bibinfo
  {author} {\bibfnamefont {S.}~\bibnamefont {Yan}}, \bibinfo {author}
  {\bibfnamefont {S.}~\bibnamefont {Kang}},\ and\ \bibinfo {author}
  {\bibfnamefont {L.}~\bibnamefont {Mei}},\ }\bibfield  {title} {\bibinfo
  {title} {Electrical field enhanced interfacial {D}zyaloshinskii-{M}oriya
  interaction in {MgO}/{Fe}/{Pt} system},\ }\href@noop {} {\bibfield  {journal}
  {\bibinfo  {journal} {Appl. Phys. Lett.}\ }\textbf {\bibinfo {volume} {113}}
  (\bibinfo {year} {2018})}\BibitemShut {NoStop}%
\bibitem [{\citenamefont {Euler}(1741)}]{Euler1741}%
  \BibitemOpen
  \bibfield  {author} {\bibinfo {author} {\bibfnamefont {L.}~\bibnamefont
  {Euler}},\ }\bibfield  {title} {\bibinfo {title} {Solutio problematis ad
  geometriam situs pertinentis},\ }\href@noop {} {\bibfield  {journal}
  {\bibinfo  {journal} {Commentarii academiae scientiarum Petropolitanae}\
  }\textbf {\bibinfo {volume} {8}},\ \bibinfo {pages} {128} (\bibinfo {year}
  {1741})}\BibitemShut {NoStop}%
\bibitem [{\citenamefont {Gibbons}(1985)}]{Gibbons1985}%
  \BibitemOpen
  \bibfield  {author} {\bibinfo {author} {\bibfnamefont {A.}~\bibnamefont
  {Gibbons}},\ }\href@noop {} {\emph {\bibinfo {title} {Algorithmic graph
  theory}}}\ (\bibinfo  {publisher} {Cambridge University Press},\ \bibinfo
  {year} {Cambridge, 1985})\BibitemShut {NoStop}%
\bibitem [{\citenamefont {Fehribach}(2009)}]{Fehribach2009}%
  \BibitemOpen
  \bibfield  {author} {\bibinfo {author} {\bibfnamefont {J.~D.}\ \bibnamefont
  {Fehribach}},\ }\bibfield  {title} {\bibinfo {title} {Vector-space methods
  and {K}irchhoff graphs for reaction networks},\ }\href@noop {} {\bibfield
  {journal} {\bibinfo  {journal} {SIAM J. Appl. Math.}\ }\textbf {\bibinfo
  {volume} {70}},\ \bibinfo {pages} {543} (\bibinfo {year} {2009})}\BibitemShut
  {NoStop}%
\bibitem [{\citenamefont {Fishtik}\ \emph {et~al.}(2004)\citenamefont
  {Fishtik}, \citenamefont {Callaghan},\ and\ \citenamefont
  {Datta}}]{Fishtik2004}%
  \BibitemOpen
  \bibfield  {author} {\bibinfo {author} {\bibfnamefont {I.}~\bibnamefont
  {Fishtik}}, \bibinfo {author} {\bibfnamefont {C.~A.}\ \bibnamefont
  {Callaghan}},\ and\ \bibinfo {author} {\bibfnamefont {R.}~\bibnamefont
  {Datta}},\ }\bibfield  {title} {\bibinfo {title} {Reaction route graphs. {I}.
  {T}heory and algorithm},\ }\href {https://doi.org/10.1021/jp0374004}
  {\bibfield  {journal} {\bibinfo  {journal} {J. Phys. Chem. B}\ }\textbf
  {\bibinfo {volume} {108}},\ \bibinfo {pages} {5671} (\bibinfo {year}
  {2004})}\BibitemShut {NoStop}%
\bibitem [{\citenamefont {Sutherland}\ and\ \citenamefont
  {Shastry}(1983)}]{Sutherland1983}%
  \BibitemOpen
  \bibfield  {author} {\bibinfo {author} {\bibfnamefont {B.}~\bibnamefont
  {Sutherland}}\ and\ \bibinfo {author} {\bibfnamefont {B.~S.}\ \bibnamefont
  {Shastry}},\ }\bibfield  {title} {\bibinfo {title} {Exact solution of a large
  class of interacting quantum systems exhibiting ground state singularities},\
  }\href {https://doi.org/10.1007/BF01009806} {\bibfield  {journal} {\bibinfo
  {journal} {J. Stat. Phys.}\ }\textbf {\bibinfo {volume} {33}},\ \bibinfo
  {pages} {477} (\bibinfo {year} {1983})}\BibitemShut {NoStop}%
\bibitem [{\citenamefont {Takano}\ \emph {et~al.}(1996)\citenamefont {Takano},
  \citenamefont {Kubo},\ and\ \citenamefont {Sakamoto}}]{Takano1996}%
  \BibitemOpen
  \bibfield  {author} {\bibinfo {author} {\bibfnamefont {K.}~\bibnamefont
  {Takano}}, \bibinfo {author} {\bibfnamefont {K.}~\bibnamefont {Kubo}},\ and\
  \bibinfo {author} {\bibfnamefont {H.}~\bibnamefont {Sakamoto}},\ }\bibfield
  {title} {\bibinfo {title} {Ground states with cluster structures in a
  frustrated {H}eisenberg chain},\ }\href
  {https://doi.org/10.1088/0953-8984/8/35/009} {\bibfield  {journal} {\bibinfo
  {journal} {J. Phys.: Cond. Mat.}\ }\textbf {\bibinfo {volume} {8}},\ \bibinfo
  {pages} {6405} (\bibinfo {year} {1996})}\BibitemShut {NoStop}%
\bibitem [{\citenamefont {Ishii}\ \emph {et~al.}(2000)\citenamefont {Ishii},
  \citenamefont {Tanaka}, \citenamefont {Hori}, \citenamefont {Uekusa},
  \citenamefont {Ohashi}, \citenamefont {Tatani}, \citenamefont {Narumi},\ and\
  \citenamefont {Kindo}}]{Ishii2000}%
  \BibitemOpen
  \bibfield  {author} {\bibinfo {author} {\bibfnamefont {M.}~\bibnamefont
  {Ishii}}, \bibinfo {author} {\bibfnamefont {H.}~\bibnamefont {Tanaka}},
  \bibinfo {author} {\bibfnamefont {M.}~\bibnamefont {Hori}}, \bibinfo {author}
  {\bibfnamefont {H.}~\bibnamefont {Uekusa}}, \bibinfo {author} {\bibfnamefont
  {Y.}~\bibnamefont {Ohashi}}, \bibinfo {author} {\bibfnamefont
  {K.}~\bibnamefont {Tatani}}, \bibinfo {author} {\bibfnamefont
  {Y.}~\bibnamefont {Narumi}},\ and\ \bibinfo {author} {\bibfnamefont
  {K.}~\bibnamefont {Kindo}},\ }\bibfield  {title} {\bibinfo {title} {Gapped
  ground state in the spin-$\frac{1}{2}$ trimer chain system $\text{Cu}_3
  \text{Cl}_6 (\text{H}_2 \text{O})_2 \cdot 2 \text{H}_8 \text{C}_4
  \text{SO}_2$},\ }\href {https://doi.org/10.1143/JPSJ.69.340} {\bibfield
  {journal} {\bibinfo  {journal} {J. Phys. Soc. Jpn.}\ }\textbf {\bibinfo
  {volume} {69}},\ \bibinfo {pages} {340} (\bibinfo {year} {2000})}\BibitemShut
  {NoStop}%
\bibitem [{\citenamefont {Strečka}\ and\ \citenamefont
  {Jaščur}(2002)}]{Strecka2002}%
  \BibitemOpen
  \bibfield  {author} {\bibinfo {author} {\bibfnamefont {J.}~\bibnamefont
  {Strečka}}\ and\ \bibinfo {author} {\bibfnamefont {M.}~\bibnamefont
  {Jaščur}},\ }\bibfield  {title} {\bibinfo {title} {Unusual quantum phase in
  exactly solvable doubly decorated {I}sing-{H}eisenberg models},\ }\href
  {https://doi.org/https://doi.org/10.1002/1521-3951(200210)233:3<R12::AID-PSSB999912>3.0.CO;2-2}
  {\bibfield  {journal} {\bibinfo  {journal} {phys. stat. sol. (b)}\ }\textbf
  {\bibinfo {volume} {233}},\ \bibinfo {pages} {R12} (\bibinfo {year}
  {2002})}\BibitemShut {NoStop}%
\bibitem [{\citenamefont {Caci}\ \emph {et~al.}(2023)\citenamefont {Caci},
  \citenamefont {Karl'ov\'a}, \citenamefont {Verkholyak}, \citenamefont
  {Stre\ifmmode~\check{c}\else \v{c}\fi{}ka}, \citenamefont {Wessel},\ and\
  \citenamefont {Honecker}}]{Caci2023}%
  \BibitemOpen
  \bibfield  {author} {\bibinfo {author} {\bibfnamefont {N.}~\bibnamefont
  {Caci}}, \bibinfo {author} {\bibfnamefont {K.}~\bibnamefont {Karl'ov\'a}},
  \bibinfo {author} {\bibfnamefont {T.}~\bibnamefont {Verkholyak}}, \bibinfo
  {author} {\bibfnamefont {J.}~\bibnamefont {Stre\ifmmode~\check{c}\else
  \v{c}\fi{}ka}}, \bibinfo {author} {\bibfnamefont {S.}~\bibnamefont
  {Wessel}},\ and\ \bibinfo {author} {\bibfnamefont {A.}~\bibnamefont
  {Honecker}},\ }\bibfield  {title} {\bibinfo {title} {Phases of the
  spin-$\frac{1}{2}$ {H}eisenberg antiferromagnet on the diamond-decorated
  square lattice in a magnetic field},\ }\href
  {https://doi.org/10.1103/PhysRevB.107.115143} {\bibfield  {journal} {\bibinfo
   {journal} {Phys. Rev. B}\ }\textbf {\bibinfo {volume} {107}},\ \bibinfo
  {pages} {115143} (\bibinfo {year} {2023})}\BibitemShut {NoStop}%
\bibitem [{\citenamefont {Fradkin}(1989)}]{Fradkin1989}%
  \BibitemOpen
  \bibfield  {author} {\bibinfo {author} {\bibfnamefont {E.}~\bibnamefont
  {Fradkin}},\ }\bibfield  {title} {\bibinfo {title} {{J}ordan-{W}igner
  transformation for quantum-spin systems in two dimensions and fractional
  statistics},\ }\href {https://doi.org/10.1103/PhysRevLett.63.322} {\bibfield
  {journal} {\bibinfo  {journal} {Phys. Rev. Lett.}\ }\textbf {\bibinfo
  {volume} {63}},\ \bibinfo {pages} {322} (\bibinfo {year} {1989})}\BibitemShut
  {NoStop}%
\bibitem [{\citenamefont {von Delft}\ and\ \citenamefont
  {Schoeller}(1998)}]{Delft1998}%
  \BibitemOpen
  \bibfield  {author} {\bibinfo {author} {\bibfnamefont {J.}~\bibnamefont {von
  Delft}}\ and\ \bibinfo {author} {\bibfnamefont {H.}~\bibnamefont
  {Schoeller}},\ }\bibfield  {title} {\bibinfo {title} {Bosonization for
  beginners — refermionization for experts},\ }\href
  {https://onlinelibrary.wiley.com/doi/abs/10.1002/andp.19985100401} {\bibfield
   {journal} {\bibinfo  {journal} {Ann. Phys.}\ }\textbf {\bibinfo {volume}
  {510}},\ \bibinfo {pages} {225} (\bibinfo {year} {1998})}\BibitemShut
  {NoStop}%
\end{thebibliography}
%


\clearpage

\onecolumngrid

\begin{appendices}

\setcounter{equation}{0}
\setcounter{figure}{0} 
\setcounter{table}{0} 
\renewcommand{\thefigure}{S\arabic{figure}}
\renewcommand{\thetable}{S\Roman{table}}
\renewcommand{\theequation}{S\arabic{equation}}

In this Supplemental Material, we provide explicit derivations of the Kirchhoff rules for product eigenstates in $XXZ$ Heisenberg models with Dzyaloshinskii-Moriya interaction and local magnetic fields.
We show that the easy-axis case only permits trivial product eigenstates unless magnetic fields are applied.
We give details on the $XX$ model on a square lattice including numerical results supplementing the discussion in the main text.
Furthermore, we derive lower and upper bounds for the degeneracy of the product eigenspace.
We conclude by giving details on the degeneracy associated with quantum spin helices in spin chains.

\section{Explicit product ansatz}

\noindent
We give detailed expressions for the equations defining the product eigenstates in the main text. To this end, we consider a more general Hamiltonian than Eq.~(1) including $XXZ$ exchange interaction, DMI and on-site magnetic fields.
\begin{align}
\begin{split}
\label{eq:Hamiltonian_appendix}
	H &= \sum_{\{ i,j \} \in E} h_{ij} + \sum_{ i \in V} h_i \ \text{with} \ h_i = \mathbf{B}_i \mathbf{S}_i \ \text{and} \\
	h_{ij} &= J \left( S^x_i S^x_j + S^y_i S^y_j \right) - \kappa_{ij} D \left( S^x_i S^y_j - S^y_i S^x_j \right) + \Delta S^z_i S^z_j,
\end{split}
\end{align}
where $S^{x/y/z} = \hbar \sigma^{x/y/z}/ 2 $ with the Pauli matrices $\sigma^{x/y/z}$.
Suppose that a product state $\ket{\Psi}$ is an eigenstate of the Hamiltonian~\eqref{eq:Hamiltonian_appendix}, i.e., $H \ket{\Psi} = \varepsilon \ket{\Psi}$.
We want to derive constraining relations for the product states that are equivalent to them solving the eigenvalue equation.
To this end, we parameterize the product state in terms of angles $\vartheta_i$ and $\varphi_i$.
Any product state can be generated by local rotations 
from a reference state $\ket{\Omega} \equiv \otimes_{i \in V} \ket{\uparrow}_i$ by
\begin{align}
\begin{split}
	\label{eq:TrafoState}
	\ket{\Psi(\bm{\vartheta}, \bm{\varphi})} &= U(\bm{\vartheta}, \bm{\varphi}) \ket{\Omega} = \left( \prod_{i \in V} U_i(\vartheta_i, \varphi_i) \right) \ket{\Omega},
	\end{split}
\end{align}
where $U_i(\vartheta_i, \varphi_i) = e^{-\frac{\mathrm{i}}{\hbar} \varphi_i S_i^z} e^{-\frac{\mathrm{i}}{\hbar} \vartheta_i S_i^y}$, we abbreviate $U_i=U_i(\vartheta_i, \varphi_i)$ and $U=U(\bm{\vartheta}, \bm{\varphi})$.
In this parameterization, $\ket{\Psi(\bm{\vartheta}, \bm{\varphi})}$ being an eigenstate of $H$ is equivalent to $\ket{\Omega}$ being an eigenstate of $\tilde{H} = U^\dagger H U$.
The constraints on the angles and hence the product eigenstates are derived by acting with $\tilde{H}$ and require that $\ket{\Omega}$ is an eigenstate of $\tilde{H}$.
\begin{align}
\label{eq:TrafoHamiltonian}
	\tilde{H} &=  U^\dagger H U = \sum_{\{ i,j \} \in E} U_j^\dagger U_i^\dagger h_{ij} U_i U_j + \sum_{ i \in V} U_i^\dagger h_i U_i, \\
\begin{split}
\label{eq:Ansatz}
	\tilde{H}(\bm{\vartheta}, \bm{\varphi}) \ket{\Omega} &= \varepsilon(\bm{\vartheta}, \bm{\varphi}) \ket{\Omega} + \sum_{ i \in V} \lambda_i(\bm{\vartheta}, \bm{\varphi}) \ket{i} + \sum_{\{ i,j \} \in E} \mu_{ij}(\bm{\vartheta}, \bm{\varphi}) \ket{i, j}.
\end{split}
\end{align}
$\ket{i}$ and $\ket{j, k}$ are the states with all spins up except for a single or two spins down on vertices $i$ and $j, k$, respectively. These are the only basis states appearing after acting on the vacuum with $\tilde{H}$ due to $U$ preserving the locality of the interaction terms in the Hamiltonian.
Since $\ket{\Omega}$, $\ket{i}$ and $\ket{j, k}$ are linearly independent, $\ket{\Omega}$ is an eigenvector of $\tilde{H}$ (and $\ket{\Psi(\bm{\vartheta}, \bm{\varphi})}$ an eigenvector of $H$) if and only if
\begin{align}
\label{eq:PAE_1}
	\lambda_i(\bm{\vartheta}, \bm{\varphi}) &= 0 \quad \forall i \in V, \\
\label{eq:PAE_2}
	\mu_{ij}(\bm{\vartheta}, \bm{\varphi}) &= 0 \quad \forall \{i,j\} \in E.
\end{align}
These equations are a special case of results derived in \cite{Cerezo2015} since we impose a uniform interaction strength for all edges and a $U(1)$-symmetry, i.e., rotational symmetry about the $z$-axis, on the quadratic terms in the Hamiltonian.
We want to give explicit expressions for Eqs.~\eqref{eq:PAE_1} and~\eqref{eq:PAE_2}.
Acting on $\ket{\Omega}$, the contributions in Eq.~\eqref{eq:TrafoHamiltonian} that are quadratic in the spin operators yield
\begin{align}
\begin{split}
	\label{eq:OneBond}
	&\frac{4}{\hbar^2} U_j^\dagger U_i^\dagger h_{ij} U_i U_j \ket{\Omega} \\
	=& \left( \cos(\vartheta_i) \cos(\vartheta_j) \Delta + \sin(\vartheta_i) \sin(\vartheta_j) \left( \cos(\varphi_i - \varphi_j) J + \sin(\varphi_i - \varphi_j) \kappa_{ij} D\right) \right) \ket{\Omega} \\
	+& \left( \cos(\vartheta_i) \sin(\vartheta_j) \left( \cos(\varphi_i - \varphi_j) J + \sin(\varphi_i - \varphi_j) \kappa_{ij} D \right) - \sin(\vartheta_i) \cos(\vartheta_j) \Delta - \mathrm{i}  \sin(\vartheta_j) \left( \sin(\varphi_i - \varphi_j) J - \cos(\varphi_i - \varphi_j) \kappa_{ij} D \right) \right) \ket{i} \\
	+& \left( \cos(\vartheta_j) \sin(\vartheta_i) \left( \cos(\varphi_i - \varphi_j) J + \sin(\varphi_i - \varphi_j) \kappa_{ij} D \right) - \sin(\vartheta_j) \cos(\vartheta_i) \Delta + \mathrm{i}  \sin(\vartheta_i) \left( \sin(\varphi_i - \varphi_j) J - \cos(\varphi_i - \varphi_j) \kappa_{ij} D \right) \right) \ket{j}  \\
	+& \left( \sin(\vartheta_i) \sin(\vartheta_j) \Delta + (\cos(\vartheta_i)\cos(\vartheta_j) - 1) \left( \cos(\varphi_i - \varphi_j) J + \sin(\varphi_i - \varphi_j) \kappa_{ij} D\right)\right)\ket{i, j} \\
	+& \mathrm{i} \left((\cos(\vartheta_i) - \cos(\vartheta_j)) \left( \sin(\varphi_i - \varphi_j) J - \cos(\varphi_i - \varphi_j) \kappa_{ij} D \right) \right) \ket{i, j}.
\end{split}
\end{align}
The terms linear in the spin operators yield
\begin{align}
\begin{split}
	\label{eq:mfield}
	\frac{2}{\hbar} U_i^\dagger h_{i} U_i \ket{\Omega}
	=& \left( \sin(\vartheta_i)\cos(\varphi_i) B^x_i + \sin(\vartheta_i)\sin(\varphi_i) B^y_i + \cos(\vartheta_i) B^z_i \right) \ket{\Omega} \\
	+& \left( \cos(\vartheta_i)\cos(\varphi_i) B^x_i + \cos(\vartheta_i)\sin(\varphi_i) B^y_i - \sin(\vartheta_i) B^z_i 
	- \mathrm{i} \left( \sin(\varphi_i) B_x - \cos(\varphi_i) B_y \right)
	\right) \ket{i}. \\
\end{split}
\end{align}
As indicated in Eq.~\eqref{eq:Ansatz}, the orthonormality of $\ket{\Omega}$, $\ket{i}$ and $\ket{i, j}$ implies that $\ket{\Psi(\bm{\vartheta}, \bm{\varphi})}$ is an eigenstate if and only if the terms proportional to the $\ket{i}$ and $\ket{i, j}$ basis states vanish.
The term proportional to $\ket{\Omega}$ gives the eigenenergy
\begin{align}
\begin{split}
	\label{eq:Energy_appendix}
	\varepsilon(\bm{\vartheta}, \bm{\varphi}) =& \frac{\hbar^2}{4} \sum_{\{ i,j \} \in E} \left( \cos(\vartheta_i) \cos(\vartheta_j) \Delta + \sin(\vartheta_i) \sin(\vartheta_j) \left( \cos(\varphi_i - \varphi_j) J + \sin(\varphi_i - \varphi_j) \kappa_{ij} D\right) \right) \\
	+& \frac{\hbar}{2} \sum_{i \in V} \left( \sin(\vartheta_i)\cos(\varphi_i) B^x_i + \sin(\vartheta_i)\sin(\varphi_i) B^y_i + \cos(\vartheta_i) B^z_i \right).
\end{split}
\end{align}
The terms proportional to $\ket{i}$ yield the equations $\lambda_i(\bm{\vartheta}, \bm{\varphi}) = 0$ that eventually give rise to the vertex rule of the main text. Separated into real and imaginary part these equations read
\begin{align}
\begin{split}
	\label{eq:PAE2}
	\frac{\hbar}{2} &\sum_{j \in V : \{ i, j \} \in E} \left( \cos(\vartheta_i) \sin(\vartheta_j) \left( \cos(\varphi_i - \varphi_j) J + \sin(\varphi_i - \varphi_j) \kappa_{ij} D \right) - \sin(\vartheta_i) \cos(\vartheta_j) \Delta \right) \\
	&= - \cos(\vartheta_i)\cos(\varphi_i) B^x_i - \cos(\vartheta_i)\sin(\varphi_i) B^y_i + \sin(\vartheta_i) B^z_i,
\end{split}\\
\begin{split}
    \label{eq:PAE2_2}
	\frac{\hbar}{2} &\sum_{j \in V : \{ i, j \} \in E} \sin(\vartheta_j) \left( \sin(\varphi_i - \varphi_j) J - \cos(\varphi_i - \varphi_j) \kappa_{ij} D \right)
	= \cos(\varphi_i) B^y_i - \sin(\varphi_i) B^x_i.
\end{split}
\end{align}
The sums on the left-hand side run over the set of all nearest neighbors $j$ of the vertex $i$.
The terms proportional to $\ket{i, j}$ result in ``edge equations" containing the constraints on the relative angles between two adjacent spins.
\begin{align}
\begin{split}
	\label{eq:PAE3}
	\sin(\vartheta_i) \sin(\vartheta_j) \Delta + (\cos(\vartheta_i)\cos(\vartheta_j) - 1) \left( \cos(\varphi_i - \varphi_j) J + \sin(\varphi_i - \varphi_j) \kappa_{ij} D\right) = 0, \\ 
	(\cos(\vartheta_i) - \cos(\vartheta_j)) \left( \sin(\varphi_i - \varphi_j) J - \cos(\varphi_i - \varphi_j) \kappa_{ij} D\right) = 0.
\end{split}
\end{align}
The terms in the Hamiltonian~\eqref{eq:TrafoHamiltonian} describing the coupling to magnetic fields add terms proportional to $\ket{\Omega}$ and $\ket{i}$. This leads to a Zeeman shift $\sum_{i \in V} \langle \mathbf{S}_i \rangle_\Psi \cdot \mathbf{B}_i$ of the energies and a source term in the vertex rule but does not affect the edge equations.
In the easy-plane case discussed in the main text, including magnetic fields modifies the vertex equations~\eqref{eq:PAE2} and~\eqref{eq:PAE2_2} in the following way
\begin{align}
	\langle \dot{S^z_i} \rangle
	&= \sin(\Theta) \Bigg( \frac{\sin(\Theta)\sin(\gamma)\hbar^2 J}{4\cos(\delta)}  \sum_{j \in V : \{i, j\} \in E} \sigma_{ij}
	+ \frac{\hbar B_i^x}{2} \sin(\varphi_i) - \frac{\hbar B_i^y}{2} \cos(\varphi_i) \Bigg) = 0 \ \text{and} \\
	\left(\langle \mathbf{S}_i \rangle_\Psi \times \left( \langle \mathbf{S}_i\rangle_\Psi \times \mathbf{B}_i \right)\right)_z &= \frac{\hbar^2}{4} \sin(\Theta) \left( \cos(\Theta)\cos(\varphi_i)B^x_i + \cos(\Theta)\sin(\varphi_i)B^y_i - \sin(\Theta) B^z_i \right) = 0.
\end{align}
The energy is
\begin{equation}
    \varepsilon = \frac{\hbar^2}{4} \Delta \abs{E} + \sum_{i \in V} \langle \mathbf{S}_i \rangle_\Psi \cdot \mathbf{B}_i.
\end{equation}

\section{Easy-axis case}

\noindent
In the main text, we claim that in the easy-axis case in absence of magnetic fields, there are no other product eigenstates than the fully polarized states with all spins up or down.
This follows from deriving the constraints on the relative angles as done for the easy-plane case.
The angles of neighboring spins obey $\varphi_i-\varphi_j = \kappa_{ij} \delta$ and $\tan\left( \vartheta_i / 2 \right) / \tan\left( \vartheta_j / 2 \right) = \exp(\sigma_{ij} \mathrm{i} \gamma)$ with the sign $\sigma_{ij} = - \sigma_{ji} = \pm 1$.
In addition, the signs in the $\vartheta$-constraint sum to $\sum_{j \in V : \{i, j\} \in E} \sigma_{ij} = 0$ at each vertex $i$.
Note that, as a consistency condition, the signs in the $\vartheta$-constraint add to zero along every circuit $\Gamma \subset E$ in the graph and $\sum_{\{i, j\} \in \Gamma} \kappa_{ij} \delta \equiv 0 \ (\operatorname{mod} 2\pi)$.
The vertex rule implies that the underlying graph is an Euler graph to allow for nontrivial product eigenstates.
Any choice of an Euler orientation gives rives to an Euler circuit such that $\sigma_{ij}$ is uniform along the circuit.
This implies that the angles $\vartheta$ increase strictly monotonously along the Euler circuit since $\gamma>0$.
In particular, the relative changes in $\vartheta$ do not sum to zero violating the circuit rule for the easy-axis case.
Hence, nontrivial product eigenstates do not exist in this case.
However, including magnetic fields allows for fine-tuning of the vertex rule rule and the existence of nontrivial product eigenstates.
In this case, the energy of the product eigenstates is
\begin{equation}
\label{eq:energy_easyAxis}
    \varepsilon(\bm{\vartheta}) = \frac{\hbar^2 J}{4 \cos(\delta)} \sum_{\{ i,j \} \in E} \big( \cos(\vartheta_i) \cos(\vartheta_j) \cos(\gamma)
    + \sin(\vartheta_i) \sin(\vartheta_j)\big)
    + \sum_{i \in V} \langle \mathbf{S}_i \rangle_\Psi \cdot \mathbf{B}_i.
\end{equation}

\section{Product eigenstates in the $XX$ model on a square lattice}

\noindent In this section, we provide further information for the $XX$ model ($\gamma = \pi / 2$, $\delta=0$) on a square lattice with open boundaries. First, we determine all Kirchhoff orientations.
The Kirchhoff rules and Fig.~2a (main text) imply that the edges of the outermost plaquettes have orientations forming vortices around these plaquettes, i.e., the arrows go around the plaquette either clockwise or counterclockwise.
We claim that a choice of the plaquette orientations, clockwise or counter-clockwise, for the $N$ leftmost and $M$ bottommost plaquettes uniquely determines a valid Kirchhoff orientation.
This follows from considering a minimal square lattice, see the $(2, 2)$ case in Tab.~\ref{tab:jaegerDegen}, with three corner plaquettes having a prechosen circular orientation.
From Kirchhoff's rules, it follows by combinatorics that this uniquely determines a valid Kirchhoff orientation.
Now, the general lattice orientation can be determined by repeatedly applying the minimal lattice result starting from the bottom left corner of the lattice.
This results in $2^{N+M-1}$ different Kirchhoff orientations.

From the Kirchhoff orientations, we can generate product eigenstates. Fixing global angles $\Phi$ and $\Theta$, e.g., $\Phi = 0$ and $\Theta = \pi / 2$ for the bottommost spin on the left boundary, i.e., it pointing in the $x$-direction, these eigenstates have the following form. Choosing a vortex orientation is equivalent to choosing a direction of the spins on the boundary vertices of degree $2$. On the left boundary, they can either point in $+x$ or $-x$-direction, and on the bottom in the $+y$ or $-y$-direction.
Spins in the same row or column as the boundary spin are determined by the anti-aligned order.
Note that this is independent for each row and column.
Also, since two product states are orthogonal if and only if at one site the spins point in opposite directions, the constructed product eigenstates are all orthogonal. This gives a lower bound for the eigenspace degeneracy of $2^{N+M-1}$, which is exponential in the length of the boundary.
The total Hilbert space's dimension is $2^{N(M+1) + (N+1)M}$ and hence exponential in the area of the lattice.
The so far constructed product eigenstates do not span the full eigenspace of product eigenstates, we have yet to consider the continuous degrees of freedom $\Phi$ and $\Theta$.
This eigenspace is spanned either by varying the global angles $\Phi$ and $\Theta$ or, as described in the main text, by the $\Theta$- and $\Phi$-derivatives of the product states in a single point $(\Phi, \Theta)$.
However, the linear hull of all product eigenstates does not exhaust the total eigenspace degeneracy for states with the same eigenenergy as the product eigenstates.
There is further degeneracy not stemming from product eigenstates or derived states such as the derivatives or projections onto eigenspaces of $S^z$, see Tab.~\ref{tab:jaegerDegen}.
The same was observed for the Kagome lattice on a torus in \cite{Changlani2018}.
Nevertheless, numerics indicate that this degeneracy coincides with the appearance of product eigenstates.

\begin{table}
\begin{tabular}{|c|c|c|c|c|c|c|}
\hline
 N & M & Graph & Orientations & Product Eigenstate Degeneracy & Total Degeneracy & Hilbert Space Dim. \\
 \hline\hline
 1 & 1 & \includegraphics[scale=0.8]{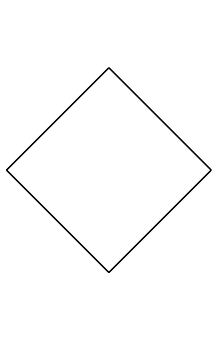} & 2 & $8$ & $10$ & 16\\
 \hline
 1 & 2 & \includegraphics[scale=0.8]{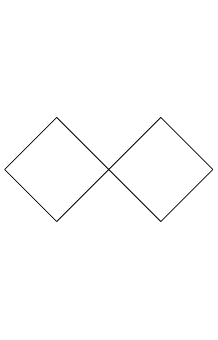} & 4 & $24$ & $44$ & 128 \\
 \hline
 1 & 3 & \includegraphics[scale=0.8]{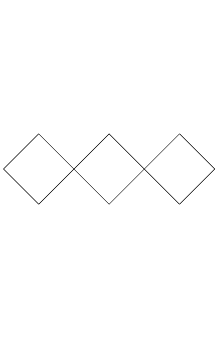} & 8 & $64$ & $234$ & 1024 \\
 \hline
 2 & 2 & \includegraphics[scale=0.8]{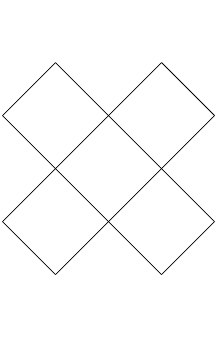} & 8 & $80$ & $280$ & 4096 \\
 \hline
\end{tabular}
\caption{Degeneracy of the subspace spanned by the product eigenstates on the square lattice ($\Delta=D=0$). $N$ is the number of rows and $M$ is the number of columns of the lattice. The number of orientations permissible with the product eigenstate equations is $2^{N+M-1}$ (fourth column) and the total Hilbert space dimension is $2^{N(M+1)+M(N+1)}$ (seventh column). The degeneracy of the product eigenstates is given in the fifth column, and furthermore, the total degeneracy of the eigenspace with energy equal to the product eigenstates' energy, the $\varepsilon$-space, is depicted in the sixth column. The latter is determined by exact diagonalization (LAPACK) neglecting energy differences below
$10^{-8}\abs{J}S^2$ ($S=\hbar /2$). \label{tab:jaegerDegen}}
\end{table}

\begin{table}
\begin{tabular}{|c|c|c|c|c|}
\hline
 N & M & Graph & Degenerate Eigenspaces at $\Delta=0$ & Degenerate Eigenspaces at $\Delta=0.1J$ \\
 \hline\hline
 1 & 1 & \includegraphics[scale=0.8]{1x1.pdf} & \includegraphics[scale=0.5]{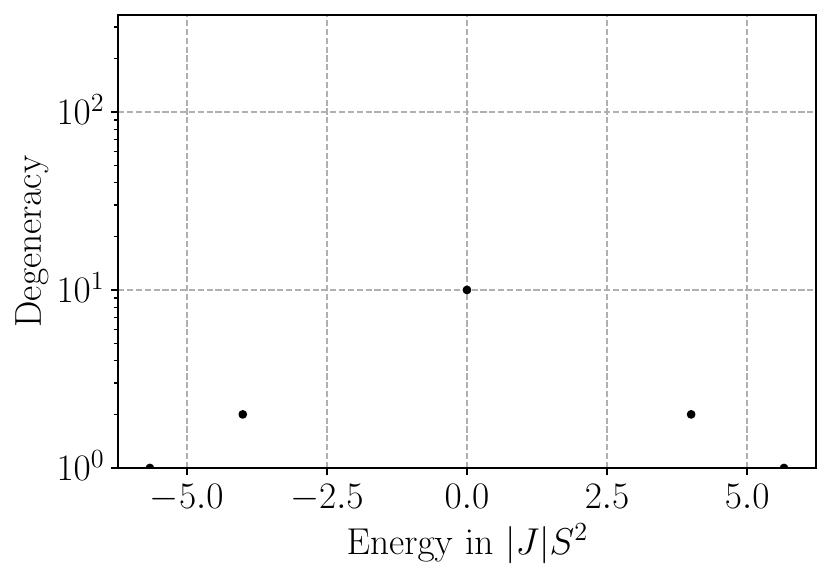} & \includegraphics[scale=0.5]{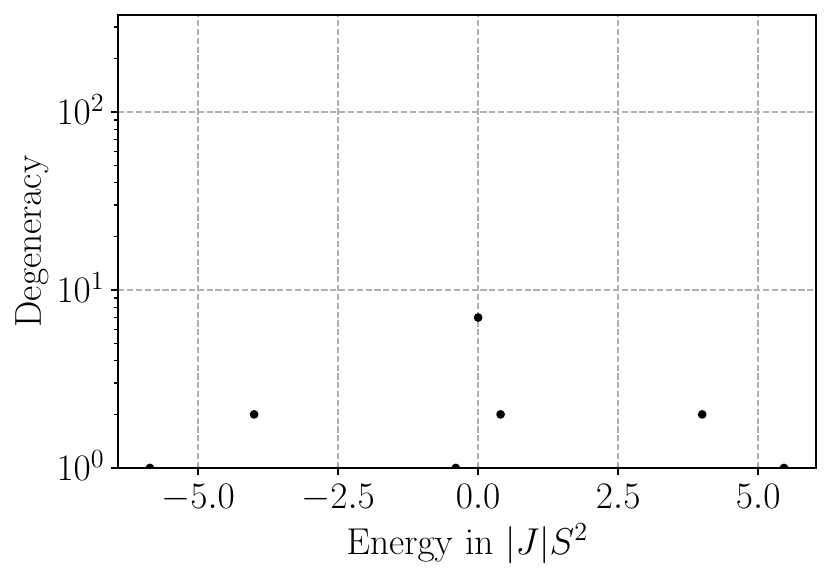} \\
 \hline
 1 & 2 & \includegraphics[scale=0.8]{1x2.pdf} & \includegraphics[scale=0.5]{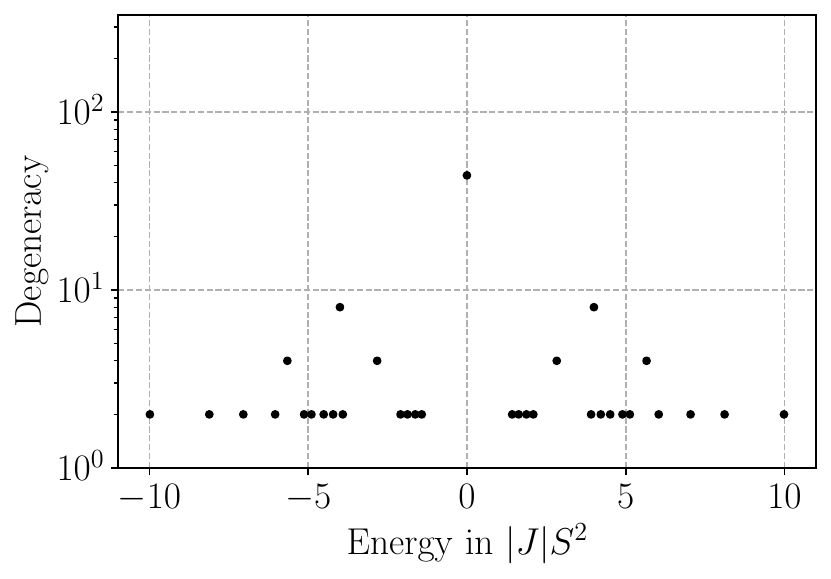} & \includegraphics[scale=0.5]{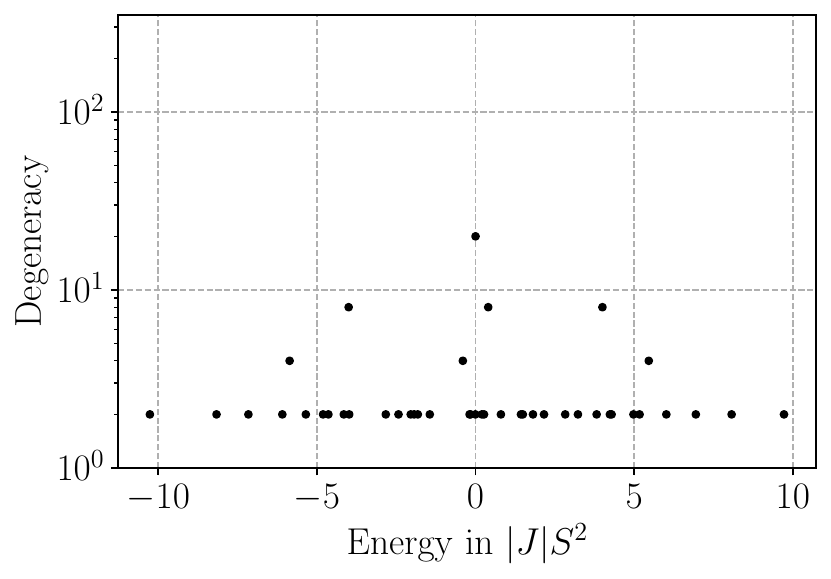} \\
 \hline
 1 & 3 & \includegraphics[scale=0.8]{1x3.pdf} & \includegraphics[scale=0.5]{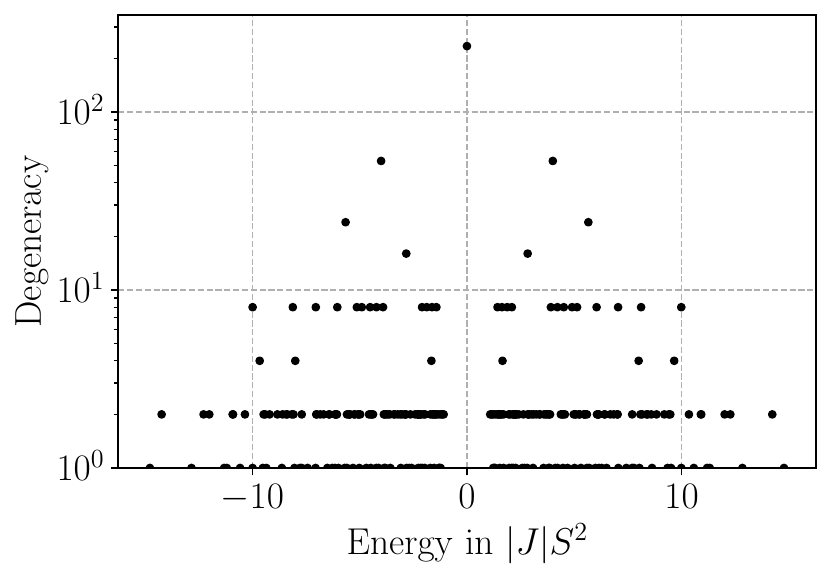} & \includegraphics[scale=0.5]{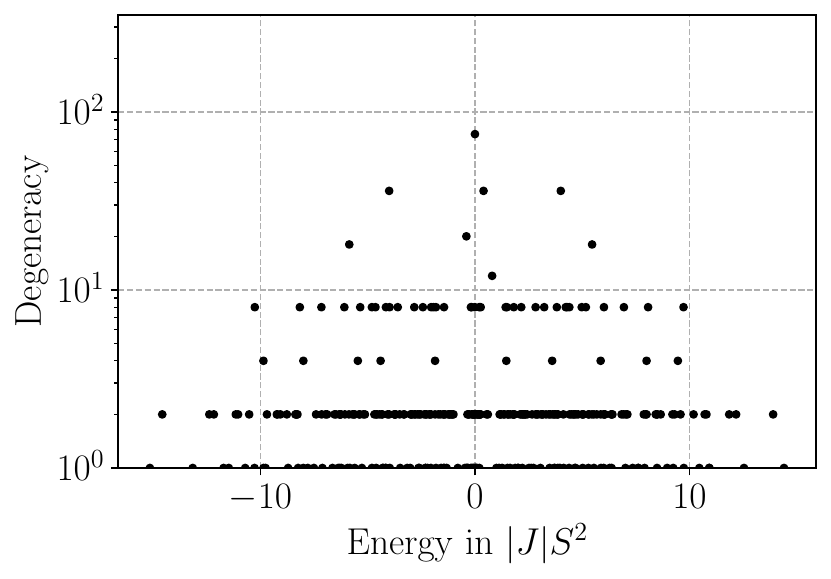} \\
 \hline
 2 & 2 & \includegraphics[scale=0.8]{2x2.pdf} & \includegraphics[scale=0.5]{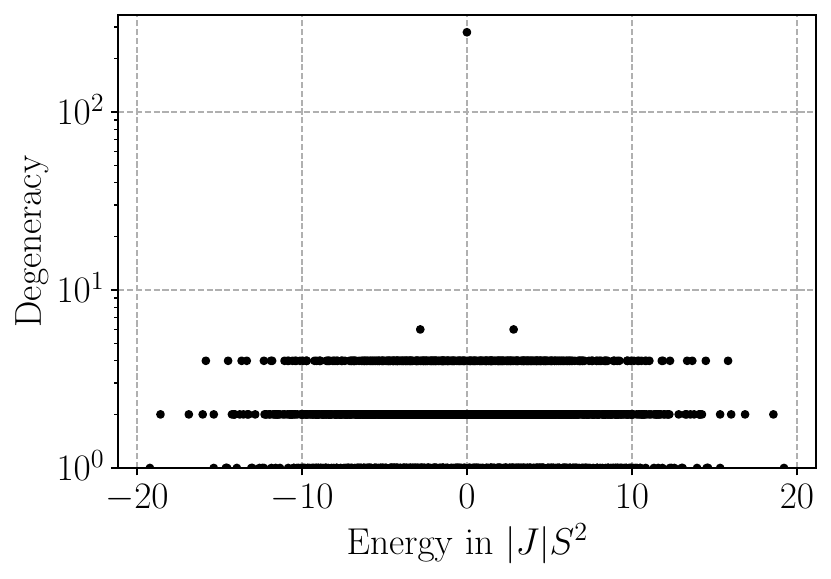} & \includegraphics[scale=0.5]{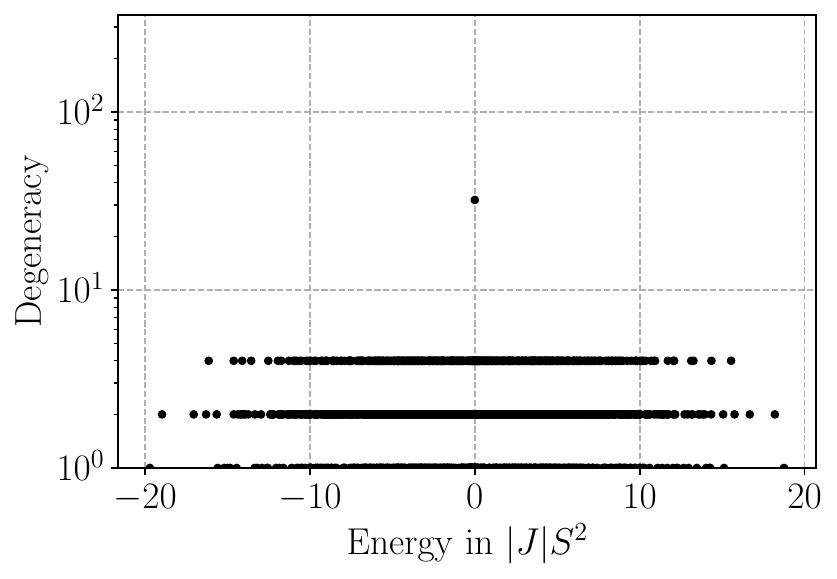} \\
 \hline
\end{tabular}
\caption{Degeneracy of the eigenspaces of the $XX$ model on the square lattice as a function of energy.
The degeneracy is determined by exact diagonalization (LAPACK) neglecting energy differences below 
$10^{-8}\abs{J}S^2$ ($S=\hbar /2$)
and depicted for two different values of the anisotropy $\Delta=0$ and $\Delta=0.1J$ ($D=0$ in both cases).
We observe a massive reduction in degeneracy of the central eigenspace that accommodates the product eigenstates for $\Delta=0$ when going to $\Delta=0.1J$.  \label{tab:jaegerDegenFigs}}
\end{table}

\section{Bounds on the degeneracy of the product eigenspace}

A lower bound for the degeneracy of the $\varepsilon$-space follows from the $U(1)$-symmetry given by rotations around the anisotropy axis. The infinitesimal generator of this symmetry operation is $S^z = \sum_{i \in V} S^z_i = \frac{\hbar}{2} \sum_{k = 0}^{\abs{V}+1} (\abs{V} - 2k) P_k$. 
Where $\abs{V}$ is the number of spins and $P_k$ the projector onto the $k^\text{th}$ eigenspace of $S^z$.
In the standard basis, these eigenspaces are spanned by the basis states with $k$ spins flipped down $\ket{l_1, \dots, l_k} = \ket{\uparrow\uparrow\uparrow\downarrow_{l_1}\uparrow \dots \uparrow\downarrow_{l_k}\uparrow}$ and $l_1 \dots l_m \in V$ denoting the positions of the spin flips in the graph.
A nontrivial product eigenstate, i.e., $\Theta \neq 0, \pi$, has non-zero weight in every eigenspace of $S^z$.
\begin{equation}
     P_k \ket{\Psi} = \exp\left(\mathrm{i}\left(k-\frac{\abs{V}}{2}\right)\Phi\right) \cos^{|V|-k} \left(\frac{\Theta}{2}\right) \sin^k \left(\frac{\Theta}{2}\right)  \sum_{l_1,\dots,l_k} \exp(\mathrm{i} Z(\sigma; l_1,…,l_k )) \ket{l_1,…,l_k},
\end{equation}
where $Z(\sigma;l_1,…,l_k )) = \frac{1}{2} \left( \sum_{j \in \{ l_1,…,l_k\}} \delta\varphi_j - \sum_{j \in V\setminus\{ l_1,…,l_k\}} \delta\varphi_j \right)$, $\delta\varphi_j = \sum_{\Gamma_{1\to j}} \left(\kappa_{ij} \delta + \sigma_{ij} \gamma\right)$ and $\Gamma_{1\to j}$ a path from the reference spin at vertex $1$ to the spin at vertex $j$.
The choice of the path is arbitrary because of the circuit rule.
From the fact that $P_k\ket{\Psi}$ only vanishes for $\Theta = 0, \pi$ and $k\neq 0,\abs{V}$, we get at least $\abs{V} + 1$ linear independent states yielding the lower bound given in the main text. Since varying $\Theta$ and $\Phi$ only changes the prefactor of $P_k\ket{\Psi}$, there are at most $\operatorname{min}\left(\abs{\sigma}, \operatorname{dim}(\operatorname{im}P_k)\right)$ linear independent states in the $k^\text{th}$ eigenspace of $S^z$ where $\operatorname{dim}(\operatorname{im}P_k) =\abs{V} \operatorname{choose} k$ is the dimension of the $k^\text{th}$ eigenspace.
This gives the upper bound on the degeneracy of the product eigenspace in the main text.

A larger lower bound for specific models can be obtained from generalizing the lower bound for the $XX$ model on a square lattice to graphs with anisotropy angles $\gamma = \pi / (2n)$, with $n$ a positive integer and an arbitrary $\delta$.
There, the number of Kirchhoff orientations gives a lower bound on the product eigenspace degeneracy.
Suppose that $\Phi = 0$ and $\Theta = \pi / 2$.
Now given two Kirchhoff orientations $\sigma$ and $\sigma^\prime$, the corresponding product eigenstates $\ket{\psi}$ and $\ket{\psi^\prime}$ are orthogonal if and only if there is a vertex $i$ such that $\varphi_i-\varphi_i^\prime \equiv \pi \ (\operatorname{mod} 2\pi)$.
This follows from the general formula for the scalar product of two product states
\begin{equation}
    \langle \Psi(\bm{\vartheta^\prime}, \bm{\varphi^\prime})  \ket{\Psi(\bm{\vartheta}, \bm{\varphi})} = \prod_{i\in V} \left( \cos\left(\frac{\vartheta_i^\prime}{2}\right) \cos\left(\frac{\vartheta_i}{2}\right) e^{\mathrm{i} \frac{\varphi_i^\prime - \varphi_i}{2}} + \sin\left(\frac{\vartheta_i^\prime}{2}\right) \sin\left(\frac{\vartheta_i}{2}\right) e^{-\mathrm{i} \frac{\varphi_i^\prime - \varphi_i}{2}} \right),
\end{equation}
which simplifies to $\langle \Psi(\bm{\vartheta^\prime}, \bm{\varphi^\prime})  \ket{\Psi(\bm{\vartheta}, \bm{\varphi})} = \prod_{i\in V} \cos((\varphi_i^\prime - \varphi_i)/2)$ for $\vartheta_i=\vartheta_i^\prime = \Theta = \pi / 2$.
Since every Kirchhoff orientation is also an Euler orientation, there exists a directed circuit $\Gamma$ that contains all edges in $E$ whose orientation is induced by $\sigma$.
From the circuit rule it follows that $\sum_{\{j,k\} \in \Gamma} \left( \sigma_{jk} - \sigma_{jk}^\prime \right)\gamma \equiv 2m\gamma \equiv 0 \ (\operatorname{mod} 2\pi)$ with $m$ the number of edges where $\sigma$ and $\sigma^\prime$ differ.
Because $\sigma \neq \sigma^\prime$, we have $m = 2n k$ for some positive integer $k$. In particular, we can find a vertex $i$ such that there are exactly $n$ edges along $\Gamma$ where $\sigma$ and $\sigma^\prime$ differ between the vertices $1$ and $i$. Then, $\varphi_i-\varphi_i^\prime = 2n \gamma = \pi$ and the assertion follows.

\section{Product eigenstates in a periodic chain}

\noindent
In this section, we give details on product eigenstates in spin chains that are helices and give the degeneracy of their eigenspaces.
The product helix solutions were derived in a number of 
previous works \cite{Cao2003,Batista2009,Cerezo2016,Popkov2021,Zhang2021-1,Zhang2021-2,Zhang2022}.
For an $XXZ$ Heisenberg quantum spin $1/2$ chain of length $N$ with periodic boundary conditions, the determining rules for the product eigenstates simplify significantly.
The circuit rule reduces to a periodic closure condition $\gamma N \equiv 0 \ (\operatorname{mod} 2\pi)$ and the change of relative $\varphi$-angle must be constant along the chain resulting in two different kinds of helices. 
The product eigenstate conditions were observed experimentally by \cite{Jepsen2022}. Their explicit form is 
\begin{equation}
    \ket{\Psi^\pm(\Theta, \Phi)} = \bigotimes_{j=1}^N \left( \cos{\left(\frac{\Theta}{2}\right)} e^{-\mathrm{i} \frac{\Phi\pm j\gamma}{2} } \ket{\uparrow}_j + \sin{\left(\frac{\Theta}{2}\right)} e^{\mathrm{i} \frac{\Phi\pm j\gamma}{2} } \ket{\downarrow}_j\right).
\end{equation}
We note that the $XXZ$ Heisenberg chain is integrable by means of the Bethe ansatz which allows, in principle, to get all eigenstates of the systems. 
The existence of these phantom helices is obscured in the standard formulation of the Bethe ansatz, where, the $U(1)$-symmetry is exploited by performing the Bethe ansatz separately in the eigenspaces of $\sum_{j=1}^N S^z_j$. The phantom helices, though, have non-zero weight in more than one eigenspace of these eigenspaces.
Phantom helices can be understood as solutions built from zero-modes acting on the pseudo-vacuum state.
They play a prominent role in the chiral reformulation of the Bethe ansatz \cite{Popkov2021,Zhang2021-1,Zhang2021-2}.
We investigated the degeneracy of the subspace spanned by the phantom helices of the $XXZ$ chain by means of exact diagonalization for chain lengths up to $N=12$, see Tab.~\ref{tab:periodicChain}.
We find that the degeneracy scales linearly in $N$. Furthermore, we find that the degeneracy due to the phantom helices generically exhausts the degeneracy of the full subspace of the same energy as the phantom helices except for few cases.

\begin{table}
\begin{tabular}{ c|c||c|c|c|c|c|c|c|c|c|c|c}
 $\Delta / J$ & $\gamma$ & $N = 2$ & $N = 3$ & $N = 4$ & $N = 5$ & $N = 6$ & $N = 7$ & $N = 8$ & $N = 9$ & $N = 10$ & $N = 11$ & $N = 12$ \\
 \hline\hline
 $1$ & $0$ & 3 & 4 & 5 & 6 & 7 & 8 & 9 & 10 & 11 & 12 & 13\\
 $-1$ & $\pi$ & 3 &  & 5 &  & 7 &  & 9 &  & 11 &  & 13 \\
 $-\frac{1}{2}$ & $\frac{2\pi}{3}$ &  & 6 &  &  & 12 &  &  & $18  (24)$ &  &  & $24 (48)$ \\
 $0$ & $\frac{\pi}{2}$ &  &  & $8 (10)$ &  &  &  & $16 (60)$ &  &  &  & $24 (386)$ \\
 $\frac{\sqrt{5}-1}{4}$ & $\frac{2\pi}{5}$ &  &  &  & 10 &  &  &  &  & 20 &  & \\
 $\frac{1}{2}$ & $\frac{\pi}{3}$ &  &  &  &  & 12 &  &  &  &  &  & $24 (48)$ \\
 $\cos(\frac{2\pi}{7})$ & $\frac{2\pi}{7}$ &  &  &  &  &  & 14 &  &  &  &  & \\
 $\frac{1}{\sqrt{2}}$ & $\frac{\pi}{4}$ &  &  &  &  &  &  & 16 &  & &  & \\
 $\cos(\frac{2\pi}{9})$ & $\frac{2\pi}{9}$ &  &  &  &  &  &  &  & 18 &  &  & \\
 $\frac{\sqrt{5}+1}{4}$ & $\frac{\pi}{5}$ &  &  &  &  &  &  &  &  & 20 &  & \\
 $\cos(\frac{2\pi}{11})$ & $\frac{2\pi}{11}$ &  &  &  &  &  &  &  &  & & 22 &\\
 $\frac{\sqrt{3}}{2}$ & $\frac{\pi}{6}$ &  &  &  &  &  &  &  &  &  &  & 24
\end{tabular}
\caption{Dimension of the subspace spanned by product eigenstates of the linear spin chain with periodic boundary conditions for different critical values of the anisotropy angle $\gamma$ and $\delta=0$. The results are obtained from exact diagonalization (LAPACK). The dimension of the total subspace of all eigenstates of the same energy as the product eigenstates is denoted in brackets if it differs from the degeneracy due to the product eigenstates. Two orthogonal eigenstates are considered to be of equal energy if their energy differs by 
$10^{-8}\abs{J}S^2$ ($S=\hbar /2$).\label{tab:periodicChain}}
\end{table}

\end{appendices}

\end{document}